\documentclass[12pt]{article}

\newlength{\abstractwidth}
\renewcommand{\thanks}[1]{\footnote{#1}} % Use this for footnotes

\makeatletter
\newcounter{fig}
\renewcommand\thefig{\arabic{fig}}

\def\fps@fig{tbp}
\def\ftype@fig{1}
\def\ext@fig{lof}
\def\fnum@fig{\figurename~\thefig}

\newenvironment{fig*}
               {\@dblfloat{fig}}
               {\end@dblfloat}

\makeatother
\usepackage{epsfig}

\newcommand{\be}{\begin{equation}}
\newcommand{\bea}{\begin{eqnarray}}
\newcommand{\eea}{\end{eqnarray}}
\newcommand{\ee}{\end{equation}}
\newcommand{\<}{\langle}
\renewcommand{\>}{\rangle}
\def\ba{\begin{eqnarray}}
\def\ea{\end{eqnarray}}

\newcommand{\ud}{{\mathrm{d}}}
\newcommand{\llr}{\longleftrightarrow}
\def\eg{{\it e.g.~}}
\def\ie{{\it i.e.~}}

\def\Tr{{\rm Tr}}

\def\half{ {1\over 2}}
\def\p{\partial}

\def\la{\label}
\def\unit{1 \hskip-.3em \raise2pt\hbox{$ \scriptstyle |$ } }
\def\ad{\stackrel{\leftrightarrow}{\partial}}
\def\aD{\stackrel{\leftrightarrow}{D}}
\def\ol{\overline}
\def\qder{\partial_{\m}^q}
\def\rder{\partial_{\n}^{\bar{r}}}
\def\psh{\partial\!\!\!\backslash}

\def\a{\alpha}

\def\d{\delta}
\def\e{\epsilon}           % Also, \varepsilon
\def\l{\lambda}
\def\m{\mu}
\def\n{\nu}
  
\def\q{\theta}                    %     \vartheta
\def\O{\Omega}

\def\O{{\cal O}}

\def\bop#1{\setbox0=\hbox{$#1M$}\mkern1.5mu
        \vbox{\hrule height0pt depth.04\ht0
        \hbox{\vrule width.04\ht0 height.9\ht0 \kern.9\ht0
        \vrule width.04\ht0}\hrule height.04\ht0}\mkern1.5mu}
\def\Box{{\mathpalette\bop{}}}                        % box
\def\dg{\sp\dagger} % hermitian conjugate
\def\leftrighthookfill#1{$\mathsurround=0pt \mathord\hook#1
       \hrulefill\mathord\hook#1$}
\def\underhook#1{\vtop{\ialign{##\crcr                 % |_| under
       $\hfil\displaystyle{#1}\hfil$\crcr
       \noalign{\kern-1pt\nointerlineskip\vskip2pt}
       \leftrighthookfill5\crcr}}}

\def\to{\rightarrow}

\makeatletter

\newcommand{\quart}{\frac{1}{4}}

\begin{document}

\begin{titlepage}

\leftline{\tt hep-th/0208041}

\vskip -.8cm

\rightline{\small{\tt CTP-MIT-3296}}

\begin{center}

\vskip 1.5 cm

{\LARGE\bf Vector operators in the BMN correspondence}
\vskip .3cm
%\vskip 1.5cm
%{\large }
%\vskip 1.2cm

%\vskip 0.5cm

\vskip 1.cm

{\large Umut G{\"u}rsoy}

\vskip 0.6cm

{\it Center for Theoretical Physics, \\
Laboratory for Nuclear Science and Department of Physics, \\
Massachusetts Institute of Technology, \\
Cambridge, Massachusetts 02139, USA} \\
E-mail: {\tt umut@mit.edu} \\

\end{center}

\vspace{1.7cm}

\begin{center}
{\bf Abstract}
\end{center}

We consider a BMN operator with one scalar, $\phi$ and one vector, $D_{\m}Z$, impurity field and compute 
the anomalous dimension both at planar and torus levels. This {\emph {mixed}} 
operator corresponds to a string state with two creation operators which belong to {\emph {different}} 
$SO(4)$ sectors of the background. 
The anomalous dimension at both levels is found to be the same as the scalar impurity BMN operator. 
At planar level this constitutes a consistency check of BMN conjecture. Agreement at the torus level can be 
explained by an argument using supersymmetry and supression in the BMN limit. The same argument implies that 
a class of fermionic BMN operators also have the same planar and 
torus level anomalous dimensions. Implications of the results for the map from ${\cal N}=4$ SYM theory to 
string theory in the pp-wave background are discussed. 

\noindent

\end{titlepage}

\newpage

\section{Introduction}

AdS/CFT correspondence, 
as an explicit realization of string/gauge duality, passed many tests performed in the
supergravity approximation over the last four 
years. Yet, this correspondence suffers, at least quantitatively, from the
obstacles in extending it into a full string theory/gauge theory
duality. This is mainly due to the lack of a clear dictionary between massive 
string modes of IIB on $AdS_5\times S^5$ and gauge invariant operators
in the dual ${\cal N}=4$ SYM at strong coupling. Specifically, the
massive modes are dual to operators in long multiplets of SYM and have
divergent anomalous dimensions as $\l=g_{YM}^2N\to\infty$. This fact, among others, 
hinders our understanding of strongly coupled gauge theory as a string theory.       

However, Berenstein, Maldacena and Nastase has taken an
important step in this direction \cite{BMN}. BMN focused on a
particular sector of the Hilbert space of gauge theory in which the
observables themselves also scale with $\l$, such that they remain
nearly BPS, namely their anomalous dimension
acquire only finite corrections. They identified the operators which 
carry large R-charge, $J$, under a $U(1)$ subgroup of $SU(4)$---
the full R-symmetry of ${\cal N}=4$ SYM---and 
this R-charge is subject to a scaling law as, $J\approx \sqrt{N}$.
As described in the next section in detail, these are essentially 
single trace operators that involve a chain of $J$ fields 
which are +1 charged under $U(1)$ with a few  
$U(1)$-neutral {\emph{impurity}} fields inserted in the chain and the number 
of these {\emph{impurities}} corresponds to the number of string excitations 
on the world-sheet.  
 
The conjecture is that, BMN operators of SYM are in one-to-one
correspondence with the string states which carry large angular
momentum, $J$, along the equator of $S^5$. The systematic way of
taking this particular limit in the gravity side is to consider 
a null geodesic along the equator and blow up the neighborhood of the geodesic
through constant scaling of the metric \cite{BFHP1}\cite{BFHP2}. 
The homogeneity property of
Einstein-Hilbert action guarantees that end-product is also a solution
of the Einstein equations, and in fact it is a plane-wave geometry
supported by the RR 5-form,   
%%%%
\be\la{plaw}
ds^2 = -4dx^+dx^- - \mu^2 z^2dx^{+2}+dz^2, \qquad
F_{+1234} = F_{+5678} = \frac{\mu}{4\pi^3g_{\rm s}\alpha^{\prime2}},
\ee
%%%%   
where $z^i$ span the 8 dimensional transverse space. 
This solution preserves 32 supercharges, just as $AdS_5\times S^5$.   
It is a particular example of the Penrose limit, \cite{Penrose}\cite{Guven} 
which generally shows that any space-time in general relativity yield plane-wave
geometry as a limit. What makes the background (\ref{plaw}) very
attractive for string theory is that quantization of string
theory in pp-wave background is known \cite{Metsaev}. RR 5-form field strength
curves the space-time in such a way that oscillator modes of the 8
transverse world sheet fields (and their fermionic partners) in the
light-cone gauge acquire a mass proportional to $F$. In turn
light-cone energy of string modes read, 
%%%%
\be\la{p-}
p^-=\mu\sum_{n=-\infty}^{\infty}N_n\sqrt{1+\frac{n^2}{(\mu
    p^+\alpha')^2}}
\ee
%%%%
where $N_n$ is the occupation number of $n$-th oscillator mode. 
In BMN dictionary this energy is dual to $\Delta-J$ of the
corresponding BMN operator in SYM, whereas the light-cone momentum
$p^+$ is proportional to the R-charge, $J$. In detail, the
correspondence is, 
%%%%
\begin{equation}\label{bmnmap}
\mu p^+\alpha' = \frac{J}{\sqrt\lambda}, \qquad
\frac{2p^-}{\mu} = \Delta-J, \qquad g_{YM}^2=4\pi g_{\rm s}.
\end{equation}
%%%%
Utilizing the AdS/CFT correspondence BMN found a relation 
between the anomalous dimensions of the BMN operators and the 
oscillation number of the corresponding string states, 
%%%%
\be\la{bmnmap2}
(\Delta-J)_{n,\,N_n} = N_n\sqrt{1+\frac{g_{YM}^2 N n^2}{J^2}}.
\ee
%%%%
We see that in the large $N$ limit, only the operators whose R-charge
scale as $J\approx \sqrt{N}$ stays in the spectrum (along with chiral
primaries) as the other observables decouple. Therefore the BMN limit 
in detail is, 
%%%%
\begin{equation}\label{bmnlimit}
N\to\infty\textrm{, with }
\frac{J}{\sqrt{N}}\textrm{ and }g_{YM}~~\textrm{ fixed.}
\end{equation}
%%%%

This limit differs from the usual large $N$ limit of gauge theory in that 
the observables are also scaled as $J$ is not fixed. Therefore neither $\l\to\infty$
implies infinite coupling in SYM nor $1/N\to 0$ implies planarity. In fact, a detailed 
study of free and coupled correlation functions in the BMN sector of SYM revealed that 
\cite{Plefka}\cite{Gross}\cite{Constable} theory develops a different 
effective coupling constant,
%%%%
\be\la{lambdaprime}
\l'=\frac{g_{YM}^2 N}{J^2}=\frac{1}{(\mu p^+\a')^2},
\ee
%%%%
and a different genus expansion parameter, 
%%%%
\be\la{g2}
g_2^2= \left(\frac{J^2}N\right)^2 = 16\pi^2 g_{\rm s}^2(\mu p^+\alpha')^4.
\ee
%%%%
As a result, in the modified large $N$ limit, (\ref{bmnlimit}), one has an 
interacting gauge theory with a tunable coupling constant $\lambda'$. However 
non-planar diagrams are not ignorable necessarily. A direct consequence of 
this non-planarity in SYM interactions can be observed as mass renormalization of string 
states\cite{Constable}. In \cite{Constable}, $\O(\l')$ contribution to the string state mass
was related to torus level contribution to $\Delta-J$ and this value was computed. 
They observed that the effective string coupling (which appears in the physical quantities like 
$\Delta-J$) is not identical to the genus counting parameter $g_2$ but modified 
with $\O(\l')$ SYM interactions as\footnote{In \cite{Rajesh}, a generalization of $g_s'$
to arbitrary values of $\lambda'$ was proposed.} 
%%%%
\be\la{effgs}
g_{s}'=g_2\sqrt{\l'}. 
\ee
%%%%     
Now, we observe a very significant fact about the BMN limit. Since $\l'$ and $g_{s}'$ 
are independent and both can be made arbitrarily small, in that regime one has {\emph{a duality between 
weakly coupled gauge theory and interacting perturbative string theory}}. This provides not only 
a duality between observables in SYM and string states on 
pp-wave background but also an explicit map between 
gauge and string interactions. 

This interacting level string/gauge duality was investigated in detail by the authors of \cite{Constable}. 
They proposed a relation between the three-string vertex and the three-point function in the gauge theory. 
At weak coupling in SYM (small $\l'$) the formula relates the matrix element of string field theory 
light-cone Hamiltonian between one-string and two-string states to the coefficient of $\O (\l')$ 
three-point function of corresponding BMN operators;
%%%%
\begin{equation}\label{7conj}
\langle i'| P^- |j'\>|k' \rangle
=\mu(\Delta_i-\Delta_j-\Delta_k)C_{ijk}
\end{equation}
%%%%
where $|i'\>$ are free string states with the normalization $\<i'|j'\>=\d_{ij}$ and $C_{ijk}$ 
is the coefficient of free planar three-point function of corresponding operators. This relation 
was recently argued to be correct in two different approaches. In the light-cone string field theory \cite{MN1} approach, 
one considers the cubic interaction term which has two constituents: 
a delta-functional overlap which is required by the continuity of world-sheet fields and 
a {\it{prefactor}} which acts on this delta-functional whose presence is required by supersymmetry. By considering 
the leading order corrections in $1/\m$ to the delta-functional, authors of 
\cite{Huang}, \cite{Durham} found agreement 
with (\ref{7conj}) in some special cases.\footnote{See also \cite{Kiem} for a test of this conjecture on supergravity modes.} 
Finally, Spradlin and Volovich \cite{MN2} obtained perfect agreement with (\ref{7conj}) 
by taking $1/\mu\to 0$ limit of the {\it prefactor}. In a totally independent approach, 
Verlinde \cite{Verlinde} developed a string bit formalism in terms of supersymmetric quantum mechanics 
for which the basic interaction also agrees with (\ref{7conj}). 

Until now all of the calculations in the literature involved BMN operators with two scalar type impurities. 
This operator corresponds to two-string excitations in the transverse directions 1,2,3 or 4 and the form 
(\ref{7conj}) was proposed only for scalar impurity BMN operators. 
However, one can generally construct BMN operators also with {\emph{vector}} or {\emph{fermionic}} type impurities. 
This manuscript is devoted to a study of the {\emph{vector}} type BMN operator which is formed 
with one scalar type impurity (a string excitation in 1,2,3,4) 
and one vector type impurity (a string excitation in 5,6,7,8). 
We study the interacting two-point function of the vector operator 
both at planar and genus one levels and compute 
the {\emph{planar}} and {\emph{torus}} level anomalous dimensions in $\O(\l')$. 
The torus dimension gives the first order, $\O({g_s'}^2)$, mass renormalization of the corresponding 
string state. Our main result is that {\emph{both}} planar {\emph{and}} torus anomalous dimensions of the 
vector operator are the same as the anomalous dimensions of the scalar BMN operator. 
   
The tensor structure of correlation functions of vector BMN operators is greatly restricted by conformal symmetry. 
In particular the unique space-time form of two-point functions is proportional to $J_{\m\n}$, the Jacobian of 
conformal inversion. 
Furthermore the vector impurity is in form of the covariant derivative $D_{\m}Z$, therefore the restrictions of 
gauge invariance on the correlators will be more apparent. 
Although the calculations are non-trivial even at the 
planar level we find that the forms required by these symmetry principles do 
emerge after combining several contributions. This fact improves our confidence in the calculations.       
    
Our motivation in studying this specific operator is two-fold. One technical motivation is to explore how 
the restrictions due to symmetry principles mentioned above are implemented. Secondly, our torus level result 
allows us to explore the implications for the interacting level string/gauge duality in the particular case of 
vector operators. 

The equality of planar level scalar and vector anomalous dimensions, $(\Delta-J)_{planar}$, is required 
by the consistency of BMN conjecture since (\ref{bmnmap2}) is independent of the transverse space 
index $i=1\dots 8$. Therefore our planar level result provides a non-trivial consistency check on the conjecture. 
However the equality at the torus level comes as a surprise at first and makes us suspect 
that there exist a superconformal transformation which relates the scalar and vector operators 
hence equates the anomalous dimensions at all genii and all orders in $\l'$. 
In section 6 we indeed find a two-step SUSY transformation which maps the scalar operator onto the vector one plus some 
correction terms. We show that these corrections terms yield an $\O(1/J)$ modification both 
for the planar and torus anomalous dimensions. 
Hence in the BMN limit these corrections are negligible and one arrives at another proof 
for the equality of scalar and vector anomalous dimensions. 
However, the fact that this SUSY map is non-exact but $1/J$ corrected prevents 
us from making a more general conclusion about the equality the dimensions 
at higher genii or higher orders in $\l'$. As a bonus, we also show that the 
fermionic type BMN operators which show up in the mentioned SUSY 
transformation possess the same planar and torus anomalous dimensions as the scalar operator.  

Our torus level result also has implications for the map between three-point functions in field theory 
and the cubic string vertex for the states with one scalar ($i=1,\dots,4$) and one vector ($i=5,\dots,8$) 
mode excited. The aforementioned {\emph{prefactor}} of string field theory weighs string interactions in the 
$i=1,\dots,4$ and $i=5,\dots,8$ directions with opposite sign, so RHS of 
(\ref{7conj}) actually vanishes in this case \cite{MN2}. A unitarity sum of the type performed for scalar excitations 
in \cite{Constable} would predict vanishing torus level anomalous dimensions in contradiction to 
our field theory result. This indicates that the anomalous dimension must come from another place in the string 
calculation, perhaps from a {\emph{contact term}}. These issues will be discussed again 
in the last section of the paper.   
 
The paper is organized as follows. In the next section we review the map between 
string states and BMN operators and introduce a neat ``$q$-variation'' notation to 
handle string excitations. Then we give the definition of the vector operator and 
compute free two-point functions at planar level. In section 3 we compute free two-point function of the 
vector operator at the torus level. This will provide a warm-up exercise for our interacting torus level 
calculations. Section 4 presents 
the calculation of planar anomalous dimension and develops necessary techniques to be
used also at torus level. In section 5 we calculate the torus dimension of the 
vector operator. Section 6 presents a SUSY argument to understand why one obtains same anomalous 
dimension for vector and scalar operators at both planar and torus levels. 
In section 7 we conclude with a summary of our results and discuss possible resolutions of 
the above mentioned contradiction between string field theory and gauge theory 
results for the torus dimension of the vector operator. 
Appendices fill in some of the details of the calculations.       
  
\section{State/operator map}

\subsection{BMN operators}

In this section we shall first review the BMN state/operator map between first few bosonic excitations 
of IIB string on the pp-wave background and corresponding 
operators in SYM on ${\mathbf{R}}\times S^3$. First of all,
string vacuum corresponds to the BPS operator (with appropriate normalization)
\footnote{We use the common convention $\Tr(T^aT^b)=\half\delta^{ab}$.},
%%%%
\be\la{vac}
O^J=\frac{1}{\sqrt{J(N/2)^J}}\Tr(Z^J) \longleftrightarrow |0,p^+\>.
\ee
%%%%
To discuss the excitations it is convenient to form complex 
combinations of  the 6 
scalars of SYM as, 
%%%%
\be\la{deffields}
Z^1 = Z = \frac{X^5+iX^6}{\sqrt{2}}, \quad 
Z^2 = \phi = \frac{X^1+iX^2}{\sqrt{2}}, \quad
Z^3 = \psi = \frac{X^3+iX^4}{\sqrt{2}}. \quad
\ee
%%%% 
Operators corresponding to the supergravity modes, $n=0$, are obtained from $O^{J+1}$ by
the action of $SO(6)$, conformal or SUSY lowering operations. For example, the particular
$SO(6)$ operation $\delta_{\phi}Z=\phi$ acting on $O^{J+1}$ yields 
(by the cyclicity of trace), 
%%%%
\be\la{sgp}
O_{\phi}^J = \frac{1}{\sqrt{J}}\delta_{\phi}O^{J+1} = \frac{1}{\sqrt{(N/2)^{J+1}}}\Tr(\phi Z^J).
\ee
%%%%
This is in correspondence with the supergravity mode $\a_0^{\phi\,\dg}|0,p^+\>$ 
where $\a_0^{\phi}=\frac{1}{\sqrt{2}}(\a^1-i\a^2)$. Similarly $\delta_{\psi}Z=\psi$ and
the translation $D_{\m}$ yields other bosonic supergravity modes, 
%%%%
\bea\la{sgmodes}
O_{\psi}^J = \frac{1}{\sqrt{(N/2)^{J+1}}}\Tr(\psi Z^J) &\llr & \a_0^{\psi\,\dg}|0,p^+\>\nonumber\\      
O_{\m}^J = \frac{1}{\sqrt{(N/2)^{J+1}}}\Tr(D_{\m}Z Z^J) &\llr & \a_0^{\m\,\dg}|0,p^+\>\la{omu}
\eea
%%%%
where $\m=5,6,7,8$ and $\a_0^{\psi}=\frac{1}{\sqrt{2}}(\a^3-i\a^4)$. 
To find the operator dual to a supergravity state with $N_0$ excitations one simply 
acts on $O^{J+N_0}$ with $N_0$ lowering operators. 

Turning now to the string excitations $n\ne 0$, we see that 
momentum conservation on the world-sheet prohibits creation of a 
single-excitation state with nonzero $n$. 
Therefore the next non-trivial string state involves two 
creation operators. Corresponding nearly BPS operators are introduced in \cite{BMN} and 
discussed in detail in the later literature but we would like to present 
here a slightly different approach with a more compact notation. 
This will prove very useful when we discuss interactions of BMN operators. 
To generalize from supergravity to string modes let us introduce a ``quantized'' 
version of the derivation rule and define a {\emph{$q$-variation}} by 
%%%%
\bea  
\delta^q(f_1(x)f_2(x)\dots f_k(x))& = & \delta^q f_1(x) f_2(x)\dots f_k(x) \,\,+\,\, 
q f_1(x)\delta^q f_2(x)\dots f_k(x) \nonumber\\ 
{} & & +\,\, \dots \,\,+ q^{k-1}f_1(x)\dots f_{k-1}(x)\delta^q f_k(x)
\la{qvar}
\eea
%%%%
where $f_i$ are arbitrary operators and $q$ is an arbitrary complex number
to be determined below.  
With this notation, 
the operator dual to single-excitation state, say $\a_n^{\phi\,\dg}|0,p^+\>$,  
can be obtained by acting on $O^{J+1}$ with $q$-variation $\delta_{\phi}^q$
with $q$ depending on $n$.  
By cyclicity of trace one gets, 
%%%%
$$\frac{1}{\sqrt{J}}\delta_{\phi}^q O^{J+1} = \frac{1}{J\sqrt{(N/2)^J}}
\left(\sum_{l=0}^{J}q^l\right)\Tr(\phi Z^J).$$
%%%%  
As mentioned above, this should vanish by momentum conservation for $n\ne 0$ 
and should reduce to (\ref{sgp}) for $n=0$. This determines $q$ at once, 
%%%%
$$
q = e^{2\pi in/(J+1)}.
$$
%%%%
Let us now determine the operator dual to the two-excitation state, 
$\a_m^{\psi\,\dg}\a_n^{\phi\,\dg}|0,p^+\>$. This is obtained by action of
$\delta_{\psi}^{q_2}\delta_{\phi}^{q_1}$ on $O^{J+2}$ with $q_1$ and $q_2$ 
depending on $n,m$ respectively. A single $q$-variation should vanish as above, 
hence fixing $q_1=e^{2\pi in/(J+2)}$, $q_2=e^{2\pi im/(J+2)}$.  
Double $q$-variation does not vanish in general because $q$-variation do not commute 
with cyclicity of trace. Furthermore in the ``dilute gas'' (large $J$) 
approximation we can neglect the case where both $\delta_{\phi}$ and $\delta_{\psi}$ 
acts on the same $Z$. Then trivial algebra gives,
%%%%
$$\frac{1}{J+2}\delta^{q_2}_{\psi}\delta^{q_1}_{\phi}O^{J+2}
= q_2 \frac{1}{(J+2)^{3/2}(N/2)^{J+2}}\left(\sum_{l=0}^{J}(q_1q_2)^l\right)\sum_{p=0}^J 
q_2^p\Tr(\phi Z^p \psi Z^{J-p}).$$
The first sum vanishes unless $q_1 q_2=1$. 
{\emph{Thus we reach at momentum conservation on the world-sheet}}, $m=-n$. Also, one 
can simply omit the phase factor $q_2$ in front since the corresponding state is defined 
only up to a phase. Therefore one gets the BMN operator with two scalar impurities, 
%%%%
\be\la{BMNop}
O_{\phi\psi}^n\equiv\frac{1}{J+2}\delta^{q_2}_{\psi}\delta^{q_1}_{\phi}O^{J+2}
= \frac{1}{\sqrt{J(N/2)^{J+2}}}\sum_{p=0}^J e^{\frac{2\pi i np}{J}}
\Tr(\phi Z^p \psi Z^{J-p})
\ee
%%%%
where we omitted $1/J$ corrections in large $J$ approximation. Without the 
``dilute gas'' approximation (for arbitrary $J$) one would get,
%%%%
\be\la{bmnopdet}
O_{\phi\psi}^n= \frac{1}{\sqrt{(J+2)(N/2)^{J+2}}}\left\{\sum_{p=0}^J e^{\frac{2\pi i np}{J+2}}
\Tr(\phi Z^p \psi Z^{J-p})+e^{-\frac{2\pi i n}{J+2}}
\Tr(\left(\delta_{\psi}\delta_{\phi}Z\right) Z^{J+1})\right\}
\ee
%%%% 
instead of (\ref{BMNop}). 
In what follows, we will refer to the specific scalar impurity operator, 
(\ref{BMNop}) as the ``BMN operator''. 
   
Generalization to $N$ string states in any transverse direction is obvious: 
Take corresponding $N$ $q_i$-variations 
(which might be $SO(6)$ transformation, translation or SUSY variation) 
of $O^{J+N}$ where $q_i$ are fixed as $q_i=e^{\frac{2\pi i n_i}{J+N}}$, 
and momentum conservation $\sum n N_n = 0$ will be automatic. 

\subsection{Vector operator and conformal invariance}

In this paper we will mainly be concerned with the vector
operator which involves two impurity fields and constructed in analogy 
with the BMN operator (\ref{BMNop}) but
with $\psi$ impurity replaced with $D_{\m}Z\,=\,\p_{\m}Z\,+\,ig
[A_{\m},Z]$. It should be defined such that it reduces to a {\emph{descendant
of the chiral primary operator}}, $\p_{\m}\Tr [Z^{J+2}]$, when the phases
are set to zero and it should be a {\emph{conformal primary}} when the phases
are present. Below, we will show that our general prescription for constructing 
BMN operators will do the job.   

Before that, let us recall that two-point function of {\emph{conformal primary}} 
vector operators $O_{\m}(x)$ and $O_{\n}(y)$ should have a specific
transformation law under conformal transformations---particularly under
inversion $x_{\m} \to \frac{x_{\m}}{x^2}$. The only possible tensorial
dependence on $x$ and $y$ can be through the determinant of inversion, 
%%%%
\be\la{jmunu}
J_{\m\n}(x-y) = \d_{\m\n}-2\frac{(x-y)_{\m}(x-y)_{\n}}{(x-y)^2}.
\ee
%%%%
Therefore the two-point function is restricted to the form,  
%%%%
$$\<O_{\m}(x)O_{\n}(y)\> \sim \frac{J_{\m\n}(x,y)}{(x-y)^{2\Delta}}$$
where $\Delta$ is the scale dimension. On the other hand, translation
descendants of scalar conformal primaries $O_{\m}(x)=\p_{\m}O(x)$, 
will have the following correlator, 
%%%%
\bea\<O_{\m}(x)O_{\n}(y)\> &\sim &
\p_{\m}\p_{\n}\frac{1}{(x-y)^{2(\Delta-1)}}\nonumber\\
{}&=&
\frac{2(\Delta-1)}{(x-y)^{2\Delta}}
\left(\delta_{\m\n}-2\Delta
\frac{(x-y)_{\m}(x-y)_{\n}}{(x-y)^2}\right).\nonumber
\eea
%%%%

We would like to see whether the vector operator 
constructed with our prescription obeys these 
restrictions. We will not assume large $J$ until 
we discuss correlators at the torus level and our construction 
will hold for {\emph{any}} $J$. Therefore our prescription
gives an analog of (\ref{bmnopdet}), 
%%%%
$$
O_{\m}^n\equiv\frac{1}{J+2}D^{q_2}_{\m}\delta^{q_1}_{\phi}O^{J+2}
=\frac{1}{\sqrt{(J+2)(N/2)^{J+2}}}\left\{\sum_{l=0}^J q_2^l\Tr\left(\phi Z^l D_{\m}Z Z^{J-l} \right)+
q_2^{J+1}\Tr\left(D_{\m}\phi Z^{J+1}\right)\right\}
$$
%%%%
where $D_{\m}^{q}$ is the gauge covariant ``$q$-derivative'' 
obeying the quantized derivation rule, (\ref{qvar}).   
For $q=1$, $q$-derivation
coincides with ordinary derivation.  We will use the following two 
forms of vector operator interchangeably, 
%%%%
\bea
O_{\m}^n &=& \frac{1}{\sqrt{(J+2)(N/2)^{J+2}}} D^q_{\m} 
\Tr\left(\phi Z^{J+1}\right)\la{vectorop}\\
{}&=&\frac{1}{\sqrt{(J+2)(N/2)^{J+2}}}\left
\{\sum_{l=0}^J q^l\Tr\left(\phi Z^l D_{\m}Z Z^{J-l} \right)+
q^{J+1}\Tr\left(D_{\m}\phi Z^{J+1}\right)\right\}\la{vectorop1}
\eea
%%%%
where, 
%%%%
\be\la{qdef}
q=e^{2\pi in/(J+2)}
\ee
%%%%

Note that in (\ref{vectorop}) the position of $\phi$ would matter generally since
the trace looses its cyclicity property under $q$-derivation. However one 
can easily check that only for the particular value (\ref{qdef}), 
the cyclicity is regained: under an arbitrary shift, say by $m$ units,  
in the position of $\phi$, $O_{\m}^n$ changes only by an overall phase $q^m$ which 
is irrelevant to physics. 

Having fixed the definition, it is now a straightforward exercise
to compute the planar, tree level contribution to the two-point function
$\<O_{\m}^n(x) \bar{O}_{\n}^m(y)\>$ directly from (\ref{vectorop1})
(or the one with $\phi$ shifted arbitrarily). Note that one can drop
the commutator term in $D_{\m}$ since it gives a $g_{YM}^2$ correction to
tree level. Denoting the scalar propagator by
$G(x,y)=\frac{1}{4\pi^2(x-y)^2}$, the result is   
%%%%
\be\la{2ptree}
\<O_{\m}^n(x)\bar{O}_{\nu}^m(y)\> =
2\d_{nm}\frac{J_{\m\n}(x-y)}{(x-y)^2}G(x,y)^{J+2}
\ee     
%%%%
for arbitrary nonzero $m,\,n$. The appearance of inversion determinant $J_{\m\n}$, 
(\ref{jmunu}) clearly shows that vector operator is a conformal primary
for arbitrary $J$, not necessarily large (at the tree level). In fact, a slight change 
in the definition of $q$ (for example $q^J=1$) would generate terms like $\O(1/J) \delta_{\m\n}$
which spoils conformal covariance for small $J$. Conformal covariance will be a helpful guide 
in the following calculations, therefore we shall stick to the definition, (\ref{qdef}).  

We also note that momentum on the world sheet is conserved at the planar level,
but we will see an explicit violation at the torus-level just like in the
case of BMN operators. At this point we want to point out an
alternative way to obtain above result directly by using
(\ref{vectorop}) with $r=e^{2\pi i m/(J+2)}$:
%%%%
\bea
\<O_{\m}^n(x)\bar{O}_{\nu}^m(y)\> &=& \frac{1}{\sqrt{(N/2)^{J+2}(J+2)}}\p^q_{\m}\p^{\bar{r}}_{\n}
\< \Tr(Z^{J+1}\phi)\Tr (\bar{Z}^{J+1}\bar{\phi}) \> \nonumber\\
{} &=& \frac{1}{J+2}\p^q_{\m}\p^{\bar{r}}_{\n}\left(\frac{1}{(x-y)^{2(J+2)}}\right)
\frac{1}{(4\pi^2)^{J+2}}\nonumber\\
{} &=& 2\d_{nm}\frac{J_{\m\n}(x-y)}{(x-y)^2}G(x,y)^{J+2}\nonumber
\eea
%%%%
using the planar tree level two-point function of chiral primaries and
the definition of $q$-derivation, (\ref{qvar}). This curious alternative
way is a consequence of the fact that $q$-derivation and contraction
operations commute with each other. 
We will use this fact to greatly simplify the calculations in
the following sections. We also note that when the phases are absent
above result trivially reduces to two-point
function of translation descendants since
$q$-derivation reduces to ordinary derivation for $q=1$. 

\section{Free Two-point Function at Torus Level}

Torus contribution to free two-point function of BMN operators was
calculated in \cite{Constable}. Analogous calculation for vector
operators is achieved simply by replacing one of the scalar
impurities, say $\psi$ field with the vector impurity $D_{\m}Z$. Since
our aim in this section is to obtain the free contribution, we can
drop the commutator term in the covariant derivative which is
$\O(g)$ and take the
impurity as $\p_{\m}Z$. Consider the generic torus diagram in fig. 1
where we show the $\phi$-line together with $\p_{\m} Z$ impurity of the
upper operator, $O_{\m}^J$, inserted at an arbitrary position and
denoted by an arrow on a $Z$-line. This arbitrary position is to be
supplied with the phase $q^l$ and summed from $l=0$ to $l=J+1$. To
obtain $\<O_{\m}^J O_{\n}^J\>_{{\mathrm{ torus}}}$ one simply takes
$\bar{r}$-derivative of this diagram. One should consider the
following two cases separately. 
%%%%%%%%%%%%%%%%%%%%%%%%%%%%%%%%%%%%%%%%%%%%%%%%%%%%%%%%%%%%%%%%%%%%%%%%%%%%%%%
\begin{figure}[htb]
\centerline{ \epsfysize=6cm\epsffile{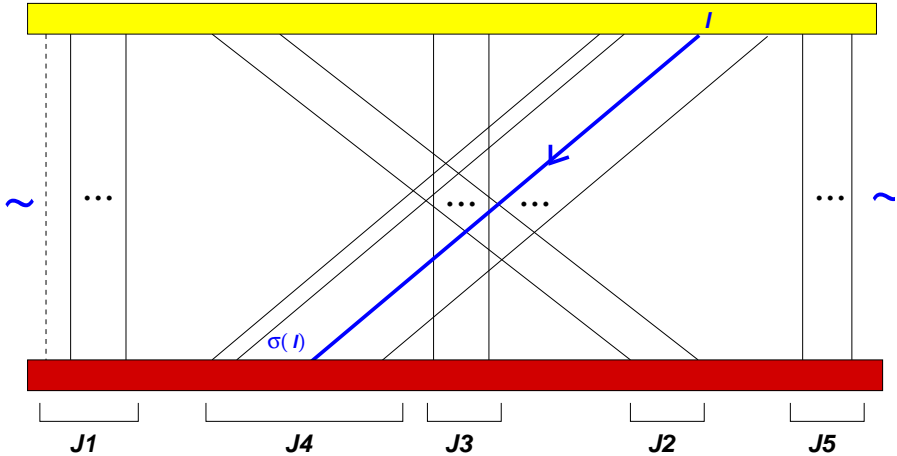}}
\caption{A typical torus digram. Dashed line represents $\phi$ and arrow on
  a solid line is $\p_{\m}Z$. The derivative $\p_{\n}$ can be placed
  on any line.}
         \label{FIG1}
\end{figure}
%%%%%%%%%%%%%%%%%%%%%%%%%%%%%%%%%%%%%%%%%%%%%%%%%%%%%%%%%%%%%%%%%%%%%%%%%%%%%%%

First consider the case when $\bar{r}$-derivative hits
the same $Z$-line as with $\p_{\m}$. Then the phase summation will
be identical to the phase sum for BMN operators, which was outlined in
section 3.3 of \cite{Constable} with a $\O(1/J)$ modification coming from
the fact that double derivative line can also coincide with the
$\phi$-line for vector operators. Here we will summarize the
calculation of \cite{Constable} for
completeness. For simplicity let us first consider operators of same momentum, \ie
$q=r$. The double derivative line may be in any of the five
groups containing  $J_1+\cdots+J_5=J+2$ lines. If it is in the first
or the last group of $J_1+J_5$ possibilities, then there is no net phase
associated with the diagram. If it is in any of the other three groups
there will be a non-trivial phase, \eg $q^{J_2+J_3}$ for the case
shown in fig. 1. Combining all possibilities the associated phase
becomes                    
$J_1 + J_2q^{J_3+J_4} + J_3q^{J_4-J_2}
+ J_4q^{-J_2-J_3}+J_5$. One should sum this phase over all
possible ways of dividing $J+2$ lines in five groups, 
$$
\frac1{(J+2)^5}
\sum_{J_1+\dots+J_5 =J+2}
\left(
J_1 + J_2q^{J_3+J_4} + J_3q^{-J_4+J_2}
+ J_4q^{J_2+J_3} + J_5
\right)$$
\bea
{} & \rightarrow_{N\to\infty} & \int_0^1 \mathrm dj_1\cdots \mathrm dj_5\,
\delta(j_1+\dots+j_5-1) \nonumber \\
{} & \times & \left(
j_1 + j_2e^{-2\pi in(j_3+j_4)} + j_3e^{2\pi in(-j_4+j_2)} +j_4e^{2\pi
in(j_2+j_3)} +j_5
\right) \nonumber \\
{} &=& \left\{ \begin{array}{ll}
\frac{1}{24}, & n=0, \\
\frac{1}{60}-\frac{1}{6(2\pi n)^2}+\frac{7}{(2\pi n)^4} & n\ne0.
\end{array}\right.\label{freegenus}
\eea
In taking the limit
$N\to\infty$ the fractions $j_i=J_i/(J+2)$ go over to continuous variables.
Apart from this phase factor there is the obvious space-time dependence

$$\frac{1}{(x-y)^{2(J+1)}} \times f_{\m\n}$$ where 
$f_{\m\n}\equiv\p_{\m}\p_{\n}\frac{1}{(x-y)^2}$.  

Now consider the second case when $\bar{q}$-derivative hits on a different
$Z$-line than $\p_{\m}$. For a fixed position of $\p_{\m}$, say $l$,
$\bar{r}$-derivative generates the phase sum 
%%%%
$$\sum_{l'=0;\,\,l'\ne
  \sigma(l)}^{J+1}\bar{q}^{l'} =  -\bar{q}^{\sigma(l)}, $$ 
%%%%
where we defined $\sigma(l)$ as the position at which $\p_{\m}Z$ connects the bottom
operator, \eg $\sigma(l)=l-(J_3+J_4)$ for the case shown in fig. 1, and used the definition
$\bar{q}^{J+2}=1$ to evaluate the sum over $l'$. Including the
summation over $l$ the total associated phase factor becomes,
%%%%%
\be\la{ps}
-\sum_{l=0}^{J+1}q^l\bar{q}^{\sigma(l)}
\ee
%%%%%
 which is obviously the same
as (\ref{freegenus}) up to a minus sign. The associated space-time
dependence is different however, 
$$\frac{1}{(x-y)^{2J}} \times f_{\m}f_{\n}$$ where $f_{\m}\equiv
  \p_{\m}\frac{1}{(x-y)^2}$. It is now easy to see that the torus
  phase factor of the vector and BMN operators will exactly be the same
  also for generic momenta $m,\,\, n$, not necessarily equal. This
  general phase factor was computed in
  \cite{Constable} and we merely quote the final result, 
%%%%
\be
A_{m,n}=\left\{\begin{array}{ll}
\frac1{24}, &  m=n=0; \\
0, & m=0, n\ne 0\,\, {\mathrm{ or }},\,\ n=0, m\ne 0; \\
\frac1{60} - \frac1{6u^2} + \frac7{u^4}, & m=n\ne 0; \\
\frac1{4u^2}\left(\frac13+\frac{35}{2u^2}\right), & m=-n\ne 0; \\
\frac1{(u-v)^2}
\left(\frac13+\frac4{v^2}+\frac4{u^2}-\frac6{uv}-\frac2{(u-v)^2}\right), 
& {\mathrm{all\,\, other\,\, cases}}
\end{array}\right.
\label{torusphase}
\ee
%%%%%
where $u=2\pi m, u=2\pi n$. Note also that space-time dependences of two
separate cases that were considered above nicely combine into the
conformal factor, (\ref{jmunu}), as 
%%%%%
\be\la{Jmunu2}\frac{f_{\m\n}}{(x-y)^2}-f_{\m}f_{\n} =
2\frac{J_{\m\n}(x,y)}{(x-y)^6}.
\ee
%%%%%
Combining above results, free two-point
function of the vector operators including genus one
corrections can now be summarized as
\begin{equation} \label{torusfree}
\langle \bar O_{\nu}^m(y)O_{\m}^n(x)\rangle_{\mathrm{free\,torus}} = 
(\delta_{nm} + g^2_2 A_{nm})\frac{2J_{\m\n}(x,y)}{(x-y)^2}G(x,y)^{J+2}.
\end{equation}

This result clearly shows the mixing of $O_{\m}^n$ operators at the torus
level since the correlator is non-zero for $n\ne m$  (unless either
$n$ or $m$ is zero). This operator mixing is described by the $\O(g_2^2)$ matrix 
$g_2^2 A_{nm}$. This particular momenta mixing issue of the BMN operators 
was first addressed in \cite{Plefka}. The eigenoperators corresponding to the 
true string eigenstates can be obtained diagonalizing the 
light cone Hamiltonian at $g_2^2$ order. We will not need this diagonalization 
explicitly for our purposes. There is another type of mixing of the BMN operators 
at the torus level: single trace operators mix with the multitrace operators at 
$\O(g_2)$ \cite{Shiraz}. Roughly, this corresponds to the fact that 
single string states are no longer the true eigenstates of the light-cone Hamiltonian 
when one considers string interactions but mixing with multi-string 
states should be taken into account. As in \cite{Constable} we shall ignore these 
mixing issues in this paper.    

\section{Planar interactions of the two-point function}

One of the main results of this manuscript is that vector
operators possess the same anomalous dimension with the BMN
operators. In this section we prove this result at the planar level
and develop the techniques necessary to handle the interactions of vector
operators which will also be used in the next section when we consider
$\O(\l')$ interactions at genus one. These techniques can easily be used
for $\O(\l')$ interactions at higher genera as well. However higher loop     
corrections would require non-trivial modifications. 

Interactions of the vector operators are far more
complicated than BMN operators because there are three new type of
interactions that has to be taken into account. Recall ${\cal N}=4$ SYM
Lagrangian (with Euclidean signature) written in  ${\cal N}=1$ component
notation \cite{D'HFS},  
%%%%
\bea
\label{lagr}
{\cal L}  &=&   \quart F_{\mu \nu }^{~~2}+\half\bar{\lambda}%
             {D}\!\!\!\!\slash\lambda
             +\overline{D_{\mu }Z^i }D_{\mu }Z^i +\half 
             \bar{\theta}^i{D}\!\!\!\!\slash\theta^i  \nonumber\\
 {}& &+ i\sqrt{2} g f^{abc}(\bar{\lambda}_{a}\bar{Z}_b^i L\theta_c^i
  -\bar{\theta}_a^i R Z_b^i\lambda_c)   
 - \frac{g}{\sqrt2} f^{abc} (\epsilon_{ijk} \bar{\theta}_a^i L Z_b^j \theta_c^k
  +\epsilon_{ijk} \bar{\theta}_a^i R \bar{Z}_b^j \theta_c^k)\nonumber    \\ 
 {}& & -\frac{1}{2} g^{2}(f^{abc} \bar{Z}_b^i Z_c^i)^{2}
  +\frac{g_{YM}^2}{2} f^{abc} f^{ade} \epsilon_{ijk} \epsilon_{ilm}
   Z_b^j Z_c^k \bar{Z}_d^l\bar{Z}_e^m  
\eea
%%%%
where $D_{\m}Z=\p_{\m}+ig[A_{\m},Z]$ and $L,R$ are the chirality
operators. For convenience we use complex combinations of the six
scalar and fermionic fields in adjoint representation,
%%%%
$$
Z^1 = Z = \frac{X^5+iX^6}{\sqrt{2}}, \quad 
Z^2 = \phi = \frac{X^1+iX^2}{\sqrt{2}}, \quad
Z^3 = \psi = \frac{X^3+iX^4}{\sqrt{2}} \quad
$$
%%%%
with analogous definitions for fermions, $\theta^i$.

Recall the result of \cite{Constable}, (also see \cite{D'HFS})
that, the only interactions involved in correlators of BMN operators
were coming from F-terms since D-term and self-energy
contributions exactly cancels each other out. That was due to a
non-renormalization theorem for two-point functions of chiral primary
operators and unfortunately, this simplification will no longer hold
for the correlators involving vector operators because of the covariant
derivatives. It will be convenient to group interactions in three main
classes because the calculation techniques that we use will differ for each separate
class: 

\begin{enumerate}
\item D-term and self energies
\item Interactions of external gluons in $O_{\m}^n$
\item F-terms
\end{enumerate}

In the following subsection we will show that interactions in the
first class can be rewritten as a correlator of the non-conserved current, 
$\<\Tr(J_{\m}\bar{J}_{\n})\>$ where 
%%%%
\be\la{current1}
J_{\m}=Z\ad_{\m}Z.
\ee
%%%%
This will be
a consequence of the non-renormalization theorem mentioned
above. Second class of interactions which are coming from the
commutator term in the covariant derivative will then promote the ordinary
derivative in $J_{\m}$ to a covariant derivative, hence total
contribution of the interactions in first and second class will be
represented as the correlator of a gauge-covariant (but
non-conserved) current $U_{\m}=Z\aD_{\m}Z$. Computation of this
correlator by differential renormalization method \cite{difrenbible}
is given in Appendix A. Last class of interactions originating from
F-term in (\ref{lagr}) are easiest to compute. This quartic vertex is
only possible between a scalar impurity, $\phi$ and an adjacent
$Z$-field (at planar level), and its contribution to the anomalous
dimension of BMN operators was already computed in
\cite{Constable}. We can confidently conclude that F-term contribution
to vector anomalous dimension is half the BMN anomalous dimension because 
the BMN operator involves two scalar impurities which contribute equally 
whereas the vector operator involves only one scalar impurity. However, a
rigorous calculation for vector operators is provided in Appendix B 
for completeness.                
    
\subsection{Non-renormalization of chiral primary correlator}

Let us begin with recalling the non-renormalization theorem of the
chiral primary correlator,
%%%%
\be\la{chiralcorr}
\<\Tr(\phi Z^{J+1})\Tr(\bar{\phi}\bar{Z}^{J+1})\>.
\ee
%%%%
For our purposes it will suffice to confine ourselves to planar
graphs. D-term part of (\ref{lagr}) includes the quartic interaction
  $\Tr([Z,\ol{Z}]^2)$ (or $\Tr([\phi,\ol{\phi}][Z,\ol{Z}])$) and
  the gluon exchange between two adjacent
  $Z$-lines, 
%%%%
\be
\label{B} \mbox{\raisebox{-.45truecm}{\epsfysize=1cm\epsffile{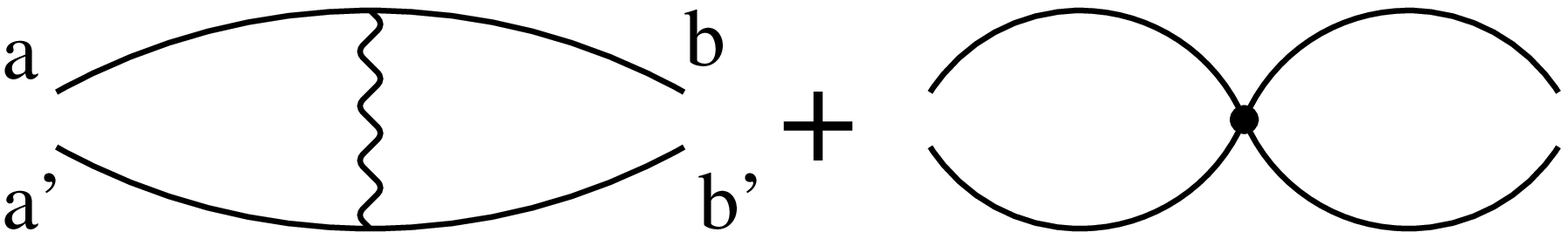} }}
 =(f^{pab} f^{p a' b'} + f^{pab'} f^{p a' b}) B(x,y) G(x,y)^2  
\ee
%%%%
where $a, a', b, b'$ indicate adjoint color indices,
$G(x,y)=1/(4\pi^2(x-y)^2) $ is the free scalar propagator and $B(x,y)$
is a function which arise from the integration over vertex positions
and contains information about the anomalous dimension. self-energy
corrections to $Z$ and $\phi$ propagators arise from a gluon exchange, chiral-chiral and
chiral-gaugino fermion loops. We represent this total self-energy contribution as, 
%%%%
\be
\label{A}
    \mbox{\raisebox{-.2truecm}{\epsfysize=0.5cm\epsffile{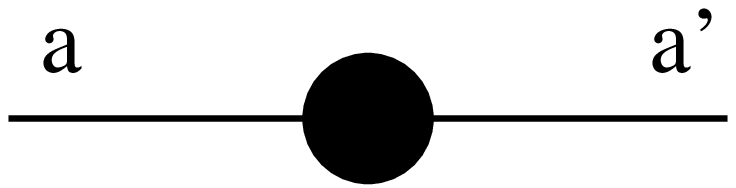} }}
    =  \delta^{aa'} N A(x,y) G(x,y) 
\ee
%%%%
where $A(x,y)$ again contains $\O(g_{YM}^2)$ contribution to anomalous
dimension. 

Planar contributions to (\ref{chiralcorr}) are obtained by inserting
(\ref{B}) in between all adjacent $Z$-$Z$ and $\phi$-$Z$
pairs and summing over self energies on all $J$ lines including
$\phi$. Since every term in (\ref{lagr}) is flavor blind except than the F-term,
eqs. (\ref{B}) and (\ref{A}) also hold for $\phi$. Therefore, from now
on we do not distinguish interactions of $\phi$ and $Z$ fields and in
all of the following figures a solid line represents either $\phi$ or $Z$ 
(unless $\phi$ is explicitly shown by a dashed line). 

A convenient way to
represent sum of all these interactions is to define a total vertex as shown
in fig. 2 and sum over $J+2$ possible insertions of this vertex in
between all adjacent lines. 
%%%%%%%%%%%%%%%%%%%%%%%%%%%%%%%%%%%%%%%%%%%%%%%%%%%%%%%%%%%%%%%%%%%%%%%%%%%%%%%
\begin{figure}[htb]
\centerline{ \epsfysize=4cm\epsffile{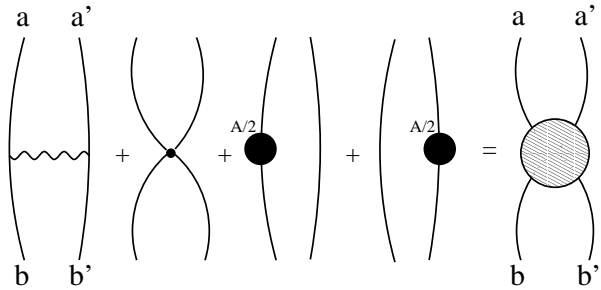}}
\caption{Combination of $g_{YM}^2$ corrections under a total vertex.}
         \label{FIG2}
\end{figure}
%%%%%%%%%%%%%%%%%%%%%%%%%%%%%%%%%%%%%%%%%%%%%%%%%%%%%%%%%%%%%%%%%%%%%%%%%%%%%%%
Note that in Fig.2 the self-energy contributions on each line are taken
as half the original value, $A(x,y)/2$, to compensate the
double-counting of self energies by this method. One of the $J+2$
possible contributions is shown in Fig.3. Using the trace identities
given in Appendix B.
 it is straightforward to compute the amplitude
represented by Fig.3. One obtains, 
%%%%
\bea
{\rm Fig.3 } &=& G(x,y)^J\left\{\half(N/2)^{J-3}\Tr(T^aT^{a'}T^{b}T^{b'})(f^{pab}
  f^{p a' b'} + f^{pab'} f^{p a' b}) B(x,y) + (N/2)^{J+1}
  A(x,y)\right\}\nonumber\\
{} &=&  G(x,y)^J (N/2)^{J+1}\left\{B(x,y)(1+\frac{2}{N^2}) +
  A(x,y)\right\}\nonumber\\
{} &\to&  G(x,y)^J (N/2)^{J+1}\left\{B(x,y) + A(x,y)\right\}\nonumber. 
\eea
$\O(1/N^2)$ term in second line is coming from the second permutation
in (\ref{B}) and is at torus order hence negligible in the BMN limit
taken in the last line. Clearly, insertions in all other spots give equal
contributions and the final answer becomes, 
%%%%
\be\la{chiralcorrl}
\<\Tr(\phi Z^{J-1})\Tr(\ol{\phi}\ol{Z}^{J-1})\> = G(x,y)^J (J+2)(N/2)^{J+1}\left\{B(x,y) + A(x,y)\right\}.
\ee
%%%%
Then, the non-renormalization theorem of this correlator
\cite{Seiberg}\cite{D'HFS} tells that,
%%%%
\be\la{nonren}
4.33 \propto\Tr\left(\mbox{\raisebox{-.15truecm}{\epsfysize=0.5cm\epsffile{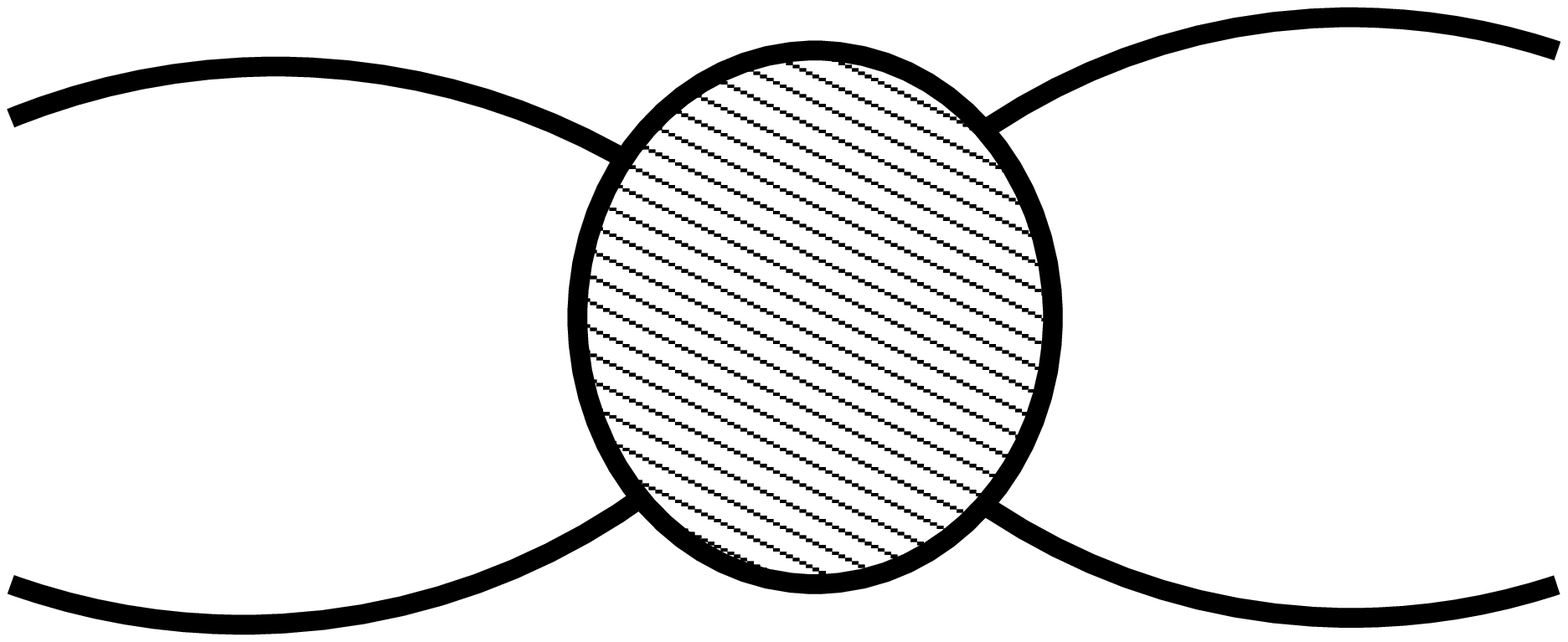}}}\right)\ 
    \propto\ (B\ +\ A) = 0 \,\,.  
\ee
%%%%
This identity greatly simplifies the following calculations. 
%%%%%%%%%%%%%%%%%%%%%%%%%%%%%%%%%%%%%%%%%%%%%%%%%%%%%%%%%%%%%%%%%%%%%%%%%%%%%%%
\begin{figure}[htb]
\centerline{ \epsfysize=6cm\epsffile{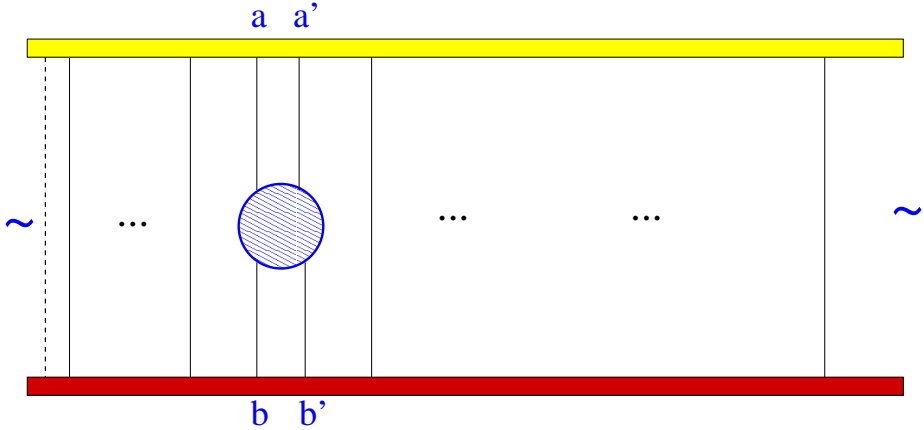}}
\caption{Planar interactions of chiral primaries can be obtained by
  placing the total vertex between all adjacent pairs. To find the
  vector correlator one simply dresses this figure by $\qder$ and $\rder$.}
         \label{FIG3}
\end{figure}
%%%%%%%%%%%%%%%%%%%%%%%%%%%%%%%%%%%%%%%%%%%%%%%%%%%%%%%%%%%%%%%%%%%%%%%%%%%%%%%

\subsection{D-term and self-energies}

Now, consider the D-term and self-energy contributions to planar
two-point function of vector operators, 
%%%%
\be\la{vectorcorr}
\<O_{\m}^n(x)\ol{O}_{\n}^m(y)\>_1=\qder\rder\<\Tr(\phi Z^{J+1})\Tr(\ol{\phi}\ol{Z}^{J+1})\>
\ee
%%%%
where we used the fact that $q$-derivation and commutation operations
commute with each other to take $q$-derivatives out of the
correlator. Once again, we note that minimal coupling in the covariant
derivative can be dropped as its contribution will be of order
$\O(g^3)$. Now it is clear that calculation is reduced to taking
$\qder\rder$ of Fig.3 and summing over all possible locations of the
total vertex in Fig.3. In taking $\qder\rder$ of Fig.3, one encounters 
three possibilities. 

If both of the derivatives hit lines other than
the four legs coming into the total vertex, than graph is proportional
to $(B+A)\p_{\m}G(x,y)\p_{\n}G(x,y)$ hence vanishes by (\ref{nonren}). Second
possibility is when one of the $q$-derivatives hit the vertex and other
outside. Supposing $\qder$ hits the vertex, trivial algebraic
manipulations show that the graph will be proportional to, 
%%%%
\bea
\p_{\n}G(x,y)\qder\Tr\left\{\mbox{\raisebox{-.15truecm}{\epsfysize=0.5cm\epsffile{2.4.eps}}}\right\}
&=&
\half\p_{\n}G(x,y)\biggl[(1-q)\Tr\left\{\mbox{\raisebox{-.2truecm}{\epsfysize=0.6cm\epsffile{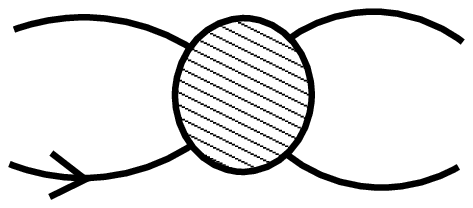}}}-
\mbox{\raisebox{-.2truecm}{\epsfysize=0.6cm\epsffile{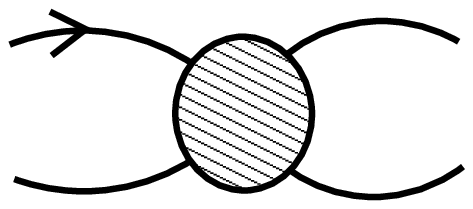}}}\right\}\nonumber\\
{}&+&(1+q)\p_{\m}\Tr\left\{\mbox{\raisebox{-.15truecm}{\epsfysize=0.5cm\epsffile{2.4.eps}}}\right\}\biggr]\nonumber\\
{}&=&
\half\p_{\n}G(x,y)\biggl[(1-q)\<\Tr(J_{\m}\bar{Z}\bar{Z})\>+(1+q)\p_{\m}\left\{(B+A)G(x,y)^2\right\}\biggr].\nonumber
\eea
%%%%
where $J_{\m}$ was defined in (\ref{current1}). The second term in the
last line again vanishes by (\ref{nonren}) whereas the first term is the self
energy and D-term corrected two point function of a vector operator
$J_{\m}$ with a scalar operator $\bar{Z}^2$. Now, it is immediate to see that by the antisymmetry of derivative in $J_{\m}$, 
both D-term quartic vertex and self ebergy corrections to $\<J_{\m}\bar{Z}^2\>$ vanishes. With a little more afford 
one can also see that the gluon exchange contribution is identically zero as well 
and the second possibility givevs no contribution.   

Therefore we are only left with the third possibility
where both $\qder$ and $\rder$ are acting on the vertex in
Fig.3. With similar algebraic manipulations of this graph and the use
of (\ref{nonren}) one obtains, 
%%%%
\be
\qder\rder\Tr\left\{\mbox{\raisebox{-.15truecm}{\epsfysize=0.5cm\epsffile{2.4.eps}}}\right\}
=\quart(1-q)(1-\bar{r})\<\Tr(J_{\m}(x)\ol{J}_{\n}(y))\>.
\ee
%%%%
Recall that there is a phase factor depending on the position of the
vertex in Fig.3. If this position is $l$ then this factor equals
$(q\bar{r})^l$ and one should sum over the vertex position from $l=0$
to $l=J+1$ to obtain the total contribution. Using our definition of
the vector phase, (\ref{qdef}), this phase summation generates
the multiplicative factor $(J+2)\delta_{mn}$. Furthermore, use of the
trace identities of Appendix B one squeezes the whole trace down to the
trace of interacting part with a multiplicative factor of $\half(N/2)^{J-1}$ 
(See Appendix B for a similar application of the trace identities). The final answer can 
be written as,
%%%%
\bea\la{D2pf}
 \<O_{\m}^n(x)\ol{O}_{\n}^m(y)\>_1&=&\qder\rder\<\Tr(\phi
 Z^{J+1})\Tr(\ol{\phi}\ol{Z}^{J+1})\>_1\nonumber\\
{}&=&G(x,y)^J\half (N/2)^{J-1}(J+2)\delta_{mn}\quart(1-q)(1-\bar{r})
\<\Tr(J_{\m}(x)\ol{J}_{\n}(y))\>.\nonumber\\
{}&&
\eea
%%%% 
Radiative corrections to this current correlator arise form three
sources: D-term quartic vertex, gluon exchange and self energies. It
is easy to see that D-term contribution vanishes identically by the
antisymmetry of $J_{\m}$ under exchange of two incoming $Z$
particles. This is shown in Appendix A. self-energy contributions are 
straightforward to calculate
with Differential Renormalization method \cite{difrenbible} and
calculations are explicitly shown in Appendix A. 

However, gluon exchange
contribution to $\<J\bar{J}\>$ is notoriously difficult to evaluate by
direct methods. Fortunately, there is the following trick
\footnote{We are grateful to Dan Freedman for pointing out this idea.}: Suppose
that one computes the true flavor-current correlator of scalar QED, 
$\<j\bar{j}\>$
with $j=\bar{Z}\ad Z$ instead. Feynman rules treat these two correlators
equivalently except than an overall minus sign ($J$ and $j$ differs 
only by replacement of a $Z$ with a $\bar{Z}$ 
then color factors at external vertices give rise to a minus sign) and 
the appropriate color factors at the vertices. Hence one can obtain
the anomalous dimension which arise from gluon exchange graph by
considering the vacuum polarization graphs of scalar QED at two-loop
order as we exlain below. There are four Feynman diagrams that 
are shown in Fig.4. Note that, the anomalous 
dimension arises from the sub-divergent pieces of the gluon exchange graph 
(when the internal vertices come close to $x$ or $y$.) 
Now, the Ward identity of scalar QED requires that the sub-divergent 
logaritmic pieces of graph I, II and III cancels each other out 
(IV do not contribute to anomalous dimension.) 
This fact allows us to compute gluon exchange in terms of I
and II which are easy enough to evaluate directly as shown in Appendix
A. When the smoke clears one obtains the total anomalous dimension
arising from D-term and self-energy part of the Lagrangian as, 
%%%%
\be\la{andim1}
 \<O_{\m}^n(x)\ol{O}_{\n}^m(y)\>_1 = -\frac{5}{8}\lambda'n^2
\delta_{mn} \log\left((x-y)^2\Lambda^2\right)\frac{J_{\m\n}(x,y)}{(x-y)^2}G(x,y)^{J+2}
\ee
%%%%
with the correct normalization of the vector operators.   
%%%%%%%%%%%%%%%%%%%%%%%%%%%%%%%%%%%%%%%%%%%%%%%%%%%%%%%%%%%%%%%%%%%%%%%%%%%%%%%
\begin{figure}[htb]
\centerline{ \epsfysize=6cm\epsffile{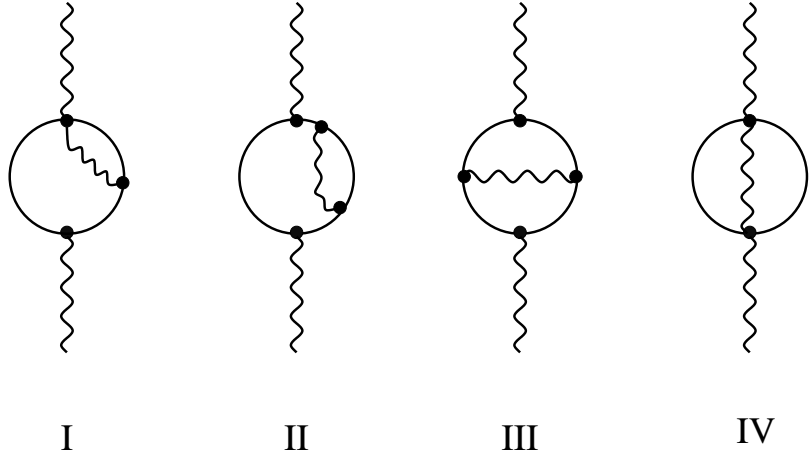}}
\caption{Two-loop diagrams of vacuum polarization in scalar
  QCD. Treatment of other four diagrams obtained by replacing the
  scalar lines with gluons can be separately and do not affect our argument.}
         \label{FIG4}
\end{figure}
%%%%%%%%%%%%%%%%%%%%%%%%%%%%%%%%%%%%%%%%%%%%%%%%%%%%%%%%%%%%%%%%%%%%%%%%%%%%%%%
\subsection{External gluons}

There are two topologically different classes of planar diagrams which
involve external gluons at $\O(g_{YM}^2)$
order. First class, shown in Fig.5, which arise from contracting
external gluons of $O_{\m}^n$ and  $\ol{O}_{\n}^m$ do not involve any
internal vertex to be integrated over, hence do not give rise to log
terms. By considering all possible Wick contractions and employing the
trace identities of Appendix B, sum over all of these diagrams yield 
(in Feynman gauge), 
%%%%
\be\la{comm1}
 \<O_{\m}^n(x)\ol{O}_{\n}^m(y)\>_2 \to 
g_{YM}^2(1-q)(1-\bar{r})\delta_{\m\n}\frac{\delta_{mn}}{(x-y)^2}(J+2)(N/2)^{J+3}G(x,y)^{J+2}. 
\ee    
%%%%
Therefore this class does not contribute to the anomalous
dimension. 

%%%%%%%%%%%%%%%%%%%%%%%%%%%%%%%%%%%%%%%%%%%%%%%%%%%%%%%%%%%%%%%%%%%%%%%%%%%%%%%
\begin{figure}[htb]
\centerline{ \epsfysize=4cm\epsffile{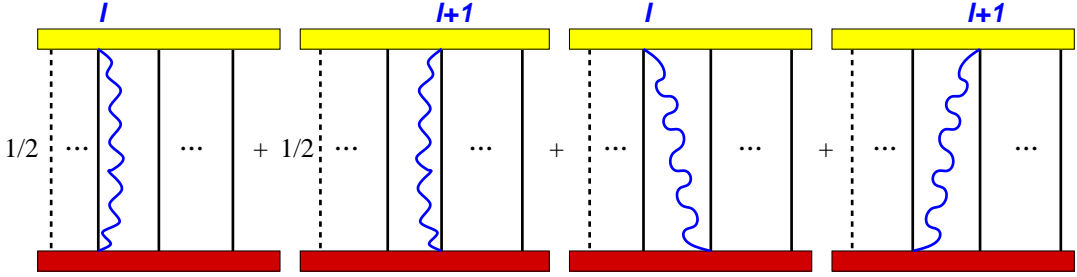}}
\caption{First class of $\O(g_{YM}^2)$ diagrams involving external
  gluons. These do not yield anomalous dimension as there are no
  internal vertices.}
         \label{FIG5}
\end{figure}
%%%%%%%%%%%%%%%%%%%%%%%%%%%%%%%%%%%%%%%%%%%%%%%%%%%%%%%%%%%%%%%%%%%%%%%%%%%%%%%
Second class of diagrams which involve one external gluon and one internal
cubic vertex are depicted in Fig.6. Diagrams where external gluon
belongs to $\ol{O}_{\n}^m$ will give identical contributions to those
in fig. 6, hence need not be considered separately. Minimal coupling in
the covariant derivative in $\ol{O}_{\n}^m$ can again be dropped since
we are interested in $\O(g_{YM}^2)$. Let us first consider graph I in
Fig.6. Total contribution to correlator is obtained by taking $\rder$
of this diagram and performing the phase sum over all possible
positions $l\,\in\,\{0,\dots,J+1\}$. Define,  
%%%%
\be\la{Cmu}\label{C}\mbox{\raisebox{-.45truecm}{\epsfysize=1cm\epsffile{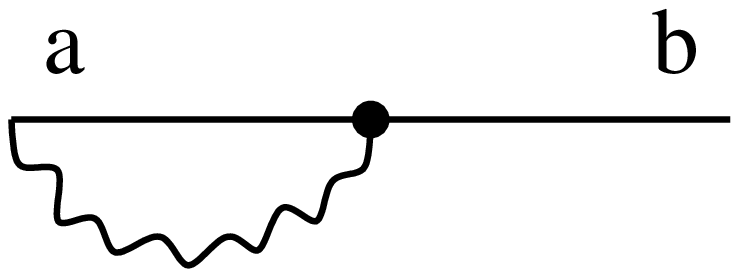}}}
 = N\delta_{ab} C_{\m}(x,y).   
\ee 
%%%%
Using the identity, $\bar{r}^{J+2}=1$ one easily obtains the $\rder$
of Graph I with the result, 
%%%%
$$(q\bar{r})^{l}G(x,y)^J\quart(N/2)^{J}N^3\left\{\p_{\n}
C_{\m}(x,y)\,G(x,y)-\p_{\n}G(x,y)C_{\m}(x,y)\right\}$$
%%%%
where we again used the trace identities of Appendix B. 
Summation over $l$ yields, 
%%%%
\be\la{exgluint}
G(x,y)^J\quart(N/2)^{J}N^3(J+2)\delta_{mn}\left\{G(x,y)\ad_{\n}C_{\m}\right\}.
\ee
%%%%    
Graph II gives identical contribution except than a factor of $q\bar{r}$. 
In Appendix A, we compute $C_{\m}(x,y)$ and conclude that graphs I and II
give the following contribution to the anomalous dimension (including
the equal contribution from the reflected graph where external gluon
belongs to $\ol{O}_{\n}^m$),
%%%%
\be\la{comm2}
{\rm Graph\ I+II} \to
-\frac{3}{8}(1+q\bar{r})\frac{g_{YM}^2N}{4\pi^2}\delta_{mn}
\frac{J_{\m\n}(x,y)}{(x-y)^2}\log\left((x-y)^2\Lambda^2\right)
G(x,y)^{J+2}.
\ee
%%%%
%%%%%%%%%%%%%%%%%%%%%%%%%%%%%%%%%%%%%%%%%%%%%%%%%%%%%%%%%%%%%%%%%%%%%%%%%%%%%%%
\begin{figure}[htb]
\centerline{ \epsfysize=4cm\epsffile{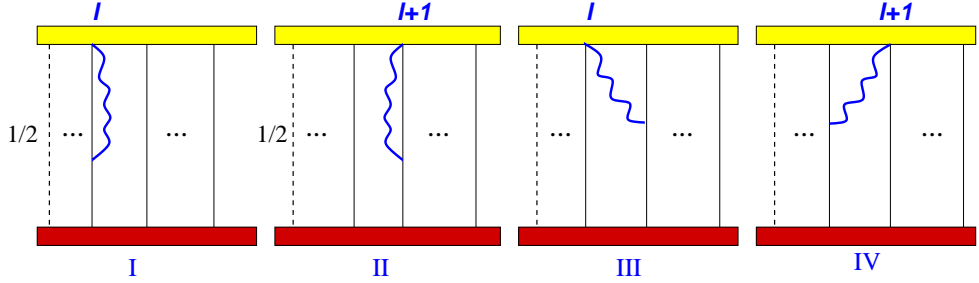}}
\caption{Second class of diagrams which involve one external
  gluon. Derivative of the $\bar{O}_{\n}^J$ can be placed at any
  position. Integration over the internal vertex yields a 
  contribution to anomalous dimension.}
         \label{FIG6}
\end{figure}
%%%%%%%%%%%%%%%%%%%%%%%%%%%%%%%%%%%%%%%%%%%%%%%%%%%%%%%%%%%%%%%%%%%%%%%%%%%%%%%

To handle graph III in Fig.6, let us write, 
%%%%
\be
\label{D} \mbox{\raisebox{-.45truecm}{\epsfysize=1cm\epsffile{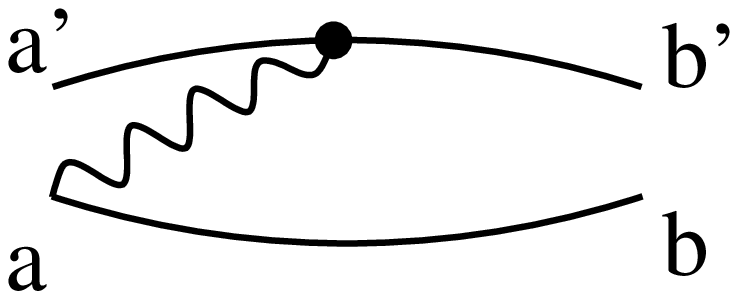}}}
 =-(f^{pab} f^{p a' b'} + f^{pab'} f^{p a' b}) C_{\m}(x,y) G(x,y).  
\ee
%%%%
By the same token, $\rder$ of graph III can be written as (at planar level), 
%%%%
\bea
-\half(N/2)^{J-1}G(x,y)^{J-1}\left(\frac{N^4}{8}\right)(q\bar{r})^l
\biggl\{\rder(G(x,y)C_{\m}(x,y))G(x,y)-\nonumber\\
-(1+\bar{r})\p_{\n}G(x,y)G(x,y)C_{\m}(x,y)\biggr\}\nonumber\\
=\half(N/2)^{J-1}G(x,y)^{J}(q\bar{r})^l\bar{r}(\frac{N^4}{8})
\left\{G(x,y)\ad_{\n}C_{\m}(x,y)\right\}
\nonumber
\eea
%%%%
where $\frac{N^4}{8}$ comes from the the color factor, $f^{pab} f^{p
  a' b'}\Tr(T^aT^{a'}T^{b'}T^b)$. Second color combination in
  (\ref{D}) gives torus level contribution hence negligible at planar level.    
Summing over $l$ and using the expression for $C_{\m}$ which is evaluated in
Appendix A, one gets the following contribution, 
%%%%
\be\la{comm3}
{\rm Graph\ III} \to
-\frac{3}{8}\frac{g_{YM}^2N}{4\pi^2}\bar{r}\delta_{mn}\frac{J_{\m\n}(x,y)}{(x-y)^2}
\log\left((x-y)^2\Lambda^2\right)
G(x,y)^{J+2}
\ee
%%%% 
where we included the equal contribution coming from the horizontal
reflection of graph IV. Graph IV and horizontal reflection of III  gives
(\ref{comm3}) with $\bar{r}$ is replaced by $q$, giving all in all, 
%%%%
\bea\la{andim2}
 \<O_{\m}^J(m;x)\ol{O}_{\n}^J(n;y)\>_2 &=& -\frac{3}{8}\frac{g_{YM}^2N}{4\pi^2}(1-q-\bar{r}+q\bar{r})
\delta_{mn}
 \log\left((x-y)^2\Lambda^2\right)\frac{J_{\m\n}(x,y)}{(x-y)^2}G(x,y)^{J+2}\nonumber\\
{}&=&-\frac{3}{8}\frac{g_{YM}^2N}{4\pi^2}(1-q)(1-\bar{r})
\delta_{mn}
 \log\left((x-y)^2\Lambda^2\right)\frac{J_{\m\n}(x,y)}{(x-y)^2}G(x,y)^{J+2}\nonumber\\
{}&=&-\frac{3}{8}\lambda'n^2
\delta_{mn} \log\left((x-y)^2\Lambda^2\right)\frac{J_{\m\n}(x,y)}{(x-y)^2}G(x,y)^{J+2}
\eea
%%%%
as the total contribution to anomalous dimension from external gluons,
after normalizing according to (\ref{vectorop}).

As an aside let us make an important observation which will be used in
section 4. In the previous section we concluded that D-term and self-energy contributions to the
vector correlator can be organized in terms of the current two-point function
$\<J_{\m}\bar{J}_{\n}\>$ where $J_{\m}=Z\ad_{\m}Z$. Curiously enough, the
external gluons result, eq. (\ref{andim2}) can exactly be reproduced (in
order $\O(g_{YM}^2)$) by promoting the ordinary derivative of $J_{\m}$
in (\ref{D2pf}) to the covariant derivative. Therefore, {\emph{one can neatly represent the contributions 
of D-term, self-energy and external gluons to the anomalous dimension in terms of radiative 
corrections to the current correlator $\<\Tr(U_{\m}(x)\ol{U}_{\n}(y))\>$}} 
where $U_{\m}$ is the {\emph{gauge-covariant}} but {\emph{non-conserved current}}, $U_{\m}=Z\aD_{\m}Z$. 
Had $U_{\m}$ been conserved there would not be any radiative corrections to the correlator 
and the corresponding non-$F$ term contributions to anomalous dimension would vanish identically. 
By using the equations of motion one can easily see that the ``non-conservation'' of $U_{\m}$ is of 
$\O(g_{YM})$ hence one expects first order corrections to $\<U\bar{U}\>$ be $\O(\l')$. 
Indeed, combining
(\ref{andim1}) and (\ref{andim2}) one obtains, 
%%%%%%%%%%%%%%%%%%%%%%%%%%%%%%%%%%%%%%%%%%%%%%%%%%%%%%%%%%%%%%%%%%%%%%%%%%%%%%%
\bea
\<O_{\m}^n(x)\ol{O}_{\n}^m(y)\>_{1+2}&=&G(x,y)^J\delta_{mn}\quart(1-q)(1-\bar{r})
\<\Tr(U_{\m}(x)\ol{U}_{\n}(y))\>\la{M2pf}\\
{}&\to&-\lambda'n^2
\delta_{mn} \log\left((x-y)^2\Lambda^2\right)\frac{J_{\m\n}(x,y)}{(x-y)^2}G(x,y)^{J+2}\la{andim3}
\eea
%%%%%%%%%%%%%%%%%%%%%%%%%%%%%%%%%%%%%%%%%%%%%%%%%%%%%%%%%%%%%%%%%%%%%%%%%%%%%%% 
Notice that anomalous dimension is in units of the correct effective `t Hooft coupling
associated with the BMN limit \ie $\lambda'\, =\  g_{YM}^2N/J^2$, and 
tensorial form of the correlator indicates that conformal primary nature
of $O_{\m}^J$ operators is preserved by planar radiative corrections.  

This result proves our previous claim that F-term contribution (which
is calculated in Appendix B) and the rest (D-term, self-energy and
external gluons) are equal hence the total planar two-point function of vector
operators with $\O(\l')$ radiative corrections can be written as, 
%%%%%%%%%%%%%%%%%%%%%%%%%%%%%%%%%%%%%%%%%%%%%%%%%%%%%%%%%%%%%%%%%%%%%%%%%%%%%%%%%%%
\be
 \<O_{\m}^n(x)\ol{O}_{\n}^m(y)\>= \left(1-\lambda'n^2\log\left((x-y)^2\Lambda^2\right)\right)
\delta_{mn}\frac{2J_{\m\n}(x,y)}{(x-y)^2}G(x,y)^{J+2}\la{andim}
\ee
%%%%%%%%%%%%%%%%%%%%%%%%%%%%%%%%%%%%%%%%%%%%%%%%%%%%%%%%%%%%%%%%%%%%%%%%%%%%%%%%%%%
which shows that {\emph{the vector operators possess the same anomalous dimension
as BMN operators, as required by the consistency of the BMN conjecture.}} 
This concludes our first test on the BMN conjecture. 

\section{Anomalous dimension on the torus} 
      
In the previous section we noted that F-term contribution to planar
anomalous dimension of vector operators is just half of the BMN case
because vector operators involve one scalar impurity field compared to 
two impurities of BMN operators. Similarly, one can easily show the effect
of F-term interactions on the torus which arise from the $\phi$ 
impurity produces half of the BMN
torus dimension. Furthermore, we will show that D-term and external
gluon contributions combine neatly into the form
$\<\Tr(U_{\m}\bar{U}_{\n})\>$ as in the planar case, hence yield the
same torus anomalous dimension as the F-terms. Therefore, the total torus
anomalous dimension of BMN and vector operators are the same as well. 
%%%%%%%%%%%%%%%%%%%%%%%%%%%%%%%%%%%%%%%%%%%%%%%%%%%%%%%%%%%%%%%%%%%%%%%%%%%%%%%
\begin{figure}[htb]
\centerline{ \epsfysize=6cm\epsffile{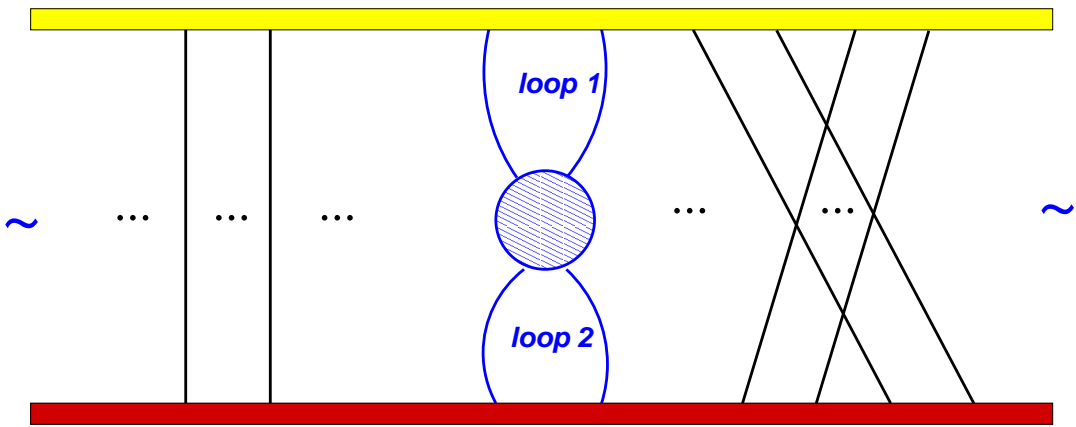}}
\caption{A generic $\O(g_{YM}^2)$ interaction on a torus diagram. 
The internal vertex generates two interaction loops in space-time graphs.} 
\label{FIG7}
\end{figure}
%%%%%%%%%%%%%%%%%%%%%%%%%%%%%%%%%%%%%%%%%%%%%%%%%%%%%%%%%%%%%%%%%%%%%%%%%%%%%%
As in the previous section we will group the interactions into D-term,
external gluon and F-terms but before that it is convenient to
classify topologically different torus diagrams which will show
up in each of these interaction
classes. Notice that all of these interactions will result in two
interaction loops which are in contact with each other at the interaction
vertex, fig. 7. Therefore, one can classify torus diagrams \cite{Constable} which
are leading order in $J$, \ie $\O(J^3)$, according to whether 
%%%%%%%%
\begin{enumerate}
\item both of the loops are contractible,
\item only one is non-contractible, 
\item both are non-contractible on the same cycle of torus,    
\item both are non-contractible on different cycles of torus.
\end{enumerate} 
%%%%%%%%
We will call these groups as {\emph{contractible}},
{\emph{semi-contractible}}, {\emph{non-contractible}} and
{\emph{special}} respectively. We called the last class
{\emph{special}} because
it is possible only for D-term
interactions as we demonstrate below. First three of these classes were discussed in 
\cite{Constable} in detail where they were called as {\emph{nearest}}, {\emph{semi-nearest}} and 
{\emph{non-nearest}} respectively. In what follows, we shall
demonstrate that only the ``non-contractible'' class gives rise to a
torus anomalous dimension. 

\subsection{Contractible diagrams}

A generic contractible diagram is displayed on the cylinder and on the periodic square
 in Figs. 8 and 9 respectively. As in planar interactions, we combined
 D-term quartic vertex with gluon exchange and
 self energies into the total vertex, see fig. 2.

%%%%%%%%%%%%%%%%%%%%%%%%%%%%%%%%%%%%%%%%%%%%%%%%%%%%%%%%%%%%%%%%%%%%%%%%%%%%%%%
\begin{figure}[htb]
\centerline{ \epsfysize=6cm\epsffile{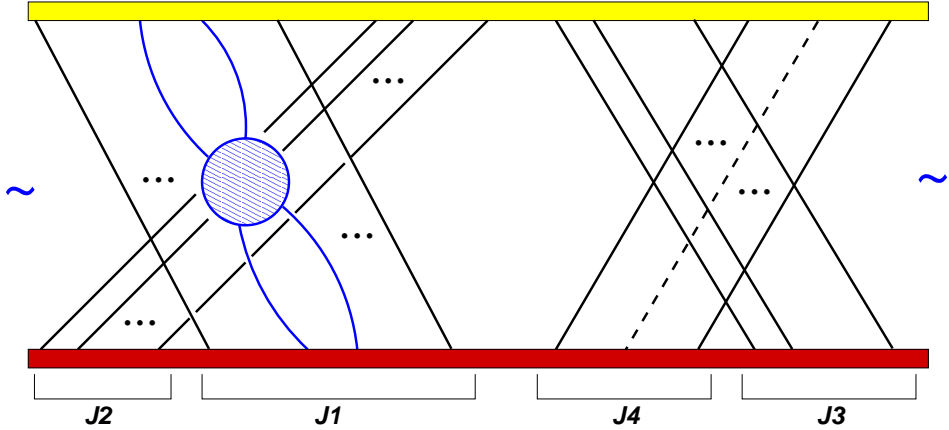}}
\caption{A generic contractible diagram. Total vertex includes, D-term quartic vertex, gluon exchange and self-energy corrections.}
\label{FIG8}
\end{figure}
%%%%%%%%%%%%%%%%%%%%%%%%%%%%%%%%%%%%%%%%%%%%%%%%%%%%%%%%%%%%%%%%%%%%%%%%%%%%%%

Let us first consider the contractible contribution to the chiral primary correlator, $\<O_{\phi}^J\bar{O}_{\phi}^J\>$ where 
$O_{\phi}^J$ is defined in (\ref{sgp}). This of course will vanish by the non-renormalization theorem of section 4.1. Still,
it will be helpful for illustrative purposes to discuss this case first because we will obtain $\<O_{\m}^n\ol{O}_{\n}^m\>$ by taking 
$q$-derivatives of the chiral primary correlator. To obtain this contribution we 
would sum the diagrams similar to Fig.8 with all possible contractible
 insertions of the this vertex. By the use of trace identities of
 Appendix B, one gets,
%%%%%%%
\be\la{chiraltorus}
\<O_{\phi}^J(x)\bar{O}_{\phi}^J(y)\> \propto (\,B(x,y)+A(x,y)\,) G(x,y)^{J+2} N^{J+1}
J^5.
\ee     
%%%%%%%
Total vertex gives the $B+A$ factor as in (\ref{chiralcorrl}), power
of $N$ indicates that this is a torus level (to be compared with
$N^{J+3}$ dependence at planar level) and the dependence on $J$ is
coming from two observations: there are $\sim J^4$ free diagrams that can be
drawn on a torus and the interaction vertex can be inserted at
$\sim J$ different positions respecting the contractibility of the
diagram. This, of course vanishes by the non-renormalization theorem,
(\ref{nonren}). One obtains correlator of vector operators simply by
taking $\qder$ and $\rder$ of fig. 8. By the same reasoning as in our
planar calculation, one sees that the only non-vanishing case occurs
when both of the derivatives hit legs of the total vertex. 
In that case one arrives at the following expression,
%%%%
$$
\frac{1}{4}q^l\bar{r}^{\sigma(l)}
(1-q)(1-\bar{r}) \<\Tr(J_{\m}\bar{J}_{\n})\> G^J N^{J-3}. 
$$     
%%%%%%%
Phase summation $\sum_{l=0}^{J+1}q^l\bar{r}^{\sigma(l)}$ is identical
to (\ref{ps}) yielding the phase factor, $A_{n,m}$,
in eq. (\ref{torusfree}). As we are interested in the anomalous
dimension, we consider the case $n=m$ and get, 
%%%%%%%
\be\la{torusD1}
\<O^n_{\m}(x)\ol{O}^n_{\n}(y)\> = \frac{A_{n,n}}{4} (1-q)(1-\bar{r}) 
\<\Tr(J_{\m}\bar{J}_{\n})\> G^J N^{J-3}.
\ee  
%%%%%%%
%%%%%%%%%%%%%%%%%%%%%%%%%%%%%%%%%%%%%%%%%%%%%%%%%%%%%%%%%%%%%%%%%%%%%%%%%%%%%%%
\begin{figure}[htb]
\centerline{ \epsfysize=6cm\epsffile{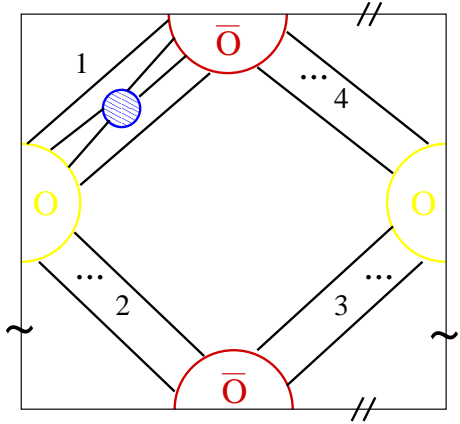}}
\caption{Same diagram as Fig.8, but represented on a periodic square.} 
\label{FIG9}
\end{figure}
%%%%%%%%%%%%%%%%%%%%%%%%%%%%%%%%%%%%%%%%%%%%%%%%%%%%%%%%%%%%%%%%%%%%%%%%%%%%%%
Apart from the phase factor associated with the topology of these
diagrams the calculation is identical to the planar case. Therefore,
together with the contribution from external gluons and F-term quartic vertex
(which is only possible between $\phi$ and adjacent $Z$'s) the
final answer can be written as, 
%%%%%%%
\be
 \<O_{\m}^n(x)\bar{O}_{\n}^n(y)\>_{contractible}= -g_2^2\lambda'n^2\log\left[(x-y)^2\Lambda^2\right]
A_{nn}\frac{2J_{\m\n}(x,y)}{(x-y)^2}G(x,y)^{J+2}.\la{torus1}
\ee
%%%%%%%
where we included the normalization associated with the torus
correlator. We conclude that contractible diagrams {\emph{do not}} 
contribute to torus anomalous dimension because their sole effect 
is to modify the normalization of the two point function by the 
factor of $A_{n,n}$. 
 
\subsection{Semi-contractible diagrams}

An example of the second class of diagrams which might potentially
contribute to torus anomalous dimension is shown on the cylinder and
the periodic square in figs. 10 and 11. However, we will now 
show that sum over all
possible semi-contractible diagrams actually vanishes. 
Last figure explicitly shows that the interaction loop which 
is formed by two adjacent $Z$ lines 
connected to $O_{\m}^n$ is contratible 
whereas the other interaction loop formed by
$\bar{Z}$ lines connected to $O_{\n}^m$ is surrounding a cycle of the
torus. A glance at either figures show that there are 8 possible
positions that one can put in such a semi-contractible vertex---as
opposed to $J$ possible insertions of contractible vertex---hence
the multiplicity of this class of graphs is order $J$ less than the
contractible class, that is $\O(J^4)$. As we explain
next, the phase summation provides a factor of $O(1/J)$ rendering
semi-contractible class leading order in $J$ \ie $O(J^3)$. 
%%%%%%%%%%%%%%%%%%%%%%%%%%%%%%%%%%%%%%%%%%%%%%%%%%%%%%%%%%%%%%%%%%%%%%%%%%%%%%%
\begin{figure}[htb]
\centerline{ \epsfysize=6cm\epsffile{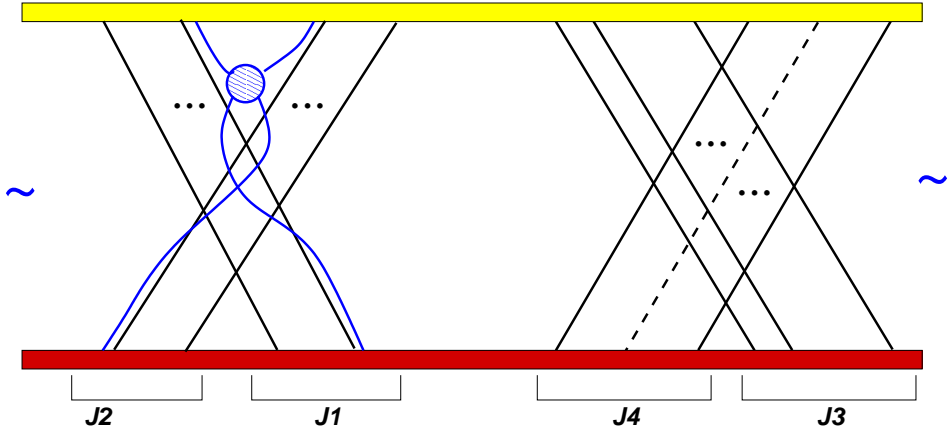}}
\caption{A semi-contractible diagram shown on the cylinder.} 
\label{FIG10}
\end{figure}
%%%%%%%%%%%%%%%%%%%%%%%%%%%%%%%%%%%%%%%%%%%%%%%%%%%%%%%%%%%%%%%%%%%%%%%%%%%%%%
To compute the contribution of fig. 10 to the total anomalous
dimension one can use the same trick as above. One first combines
D-term, gluon exchange and self energies under the total vertex of
fig.2. Then, insertion of derivatives in all possible ways with the
phases shows that,     
%%%%%
$$\<O^n_{\m}(x)\bar{O}^m_{\n}(y)\> \propto
(1-q)(1-\bar{r}^{J_1+J_2}) \<\Tr(J_{\m}\bar{J}_{\n})\>$$
%%%%%
for the fixed position of $\phi$ as in fig. 10 and fixed
$J_1\,\dots\,J_4$. Contributions from the external gluons are shown in
fig. 12. Not surprisingly, they add up with total vertex to modify the
above result as,                 
$$\<O^n_{\m}(x)\bar{O}^m_{\n}(y)\> \propto
(1-q)(1-\bar{r}^{J_1+J_2}) \<\Tr(U_{\m}\bar{U}_{\n})\>.$$
%%%%%%%%%%%%%%%%%%%%%%%%%%%%%%%%%%%%%%%%%%%%%%%%%%%%%%%%%%%%%%%%%%%%%%%%%%%%%%%
\begin{figure}[htb]
\centerline{ \epsfysize=6cm\epsffile{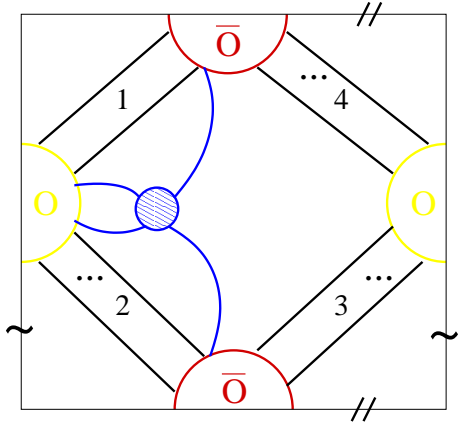}}
\caption{Same diagram as Fig.10, but represented on a periodic square.} 
\label{FIG11}
\end{figure}
%%%%%%%%%%%%%%%%%%%%%%%%%%%%%%%%%%%%%%%%%%%%%%%%%%%%%%%%%%%%%%%%%%%%%%%%%%%%%%
One similarly computes the contributions from 7 other
semi-contractible graphs with the \emph{same} $J_1\,\dots\,J_4$  
and the {\emph{fixed}} position of $\phi$, and finds that phase factors
conspire exactly to cancel out the total result. F-term contributions
also give rise to same phase factors and cancel out in exactly the
same way as described above.   
%%%%%%%%%%%%%%%%%%%%%%%%%%%%%%%%%%%%%%%%%%%%%%%%%%%%%%%%%%%%%%%%%%%%%%%%%%%%%%%
\begin{figure}[htb]
\centerline{ \epsfysize=4cm\epsffile{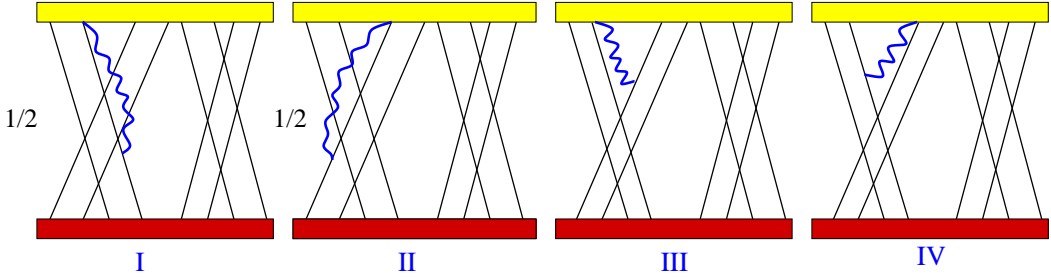}}
\caption{External gluon interactions with semi-contractible topology.} 
\label{FIG12}
\end{figure}
%%%%%%%%%%%%%%%%%%%%%%%%%%%%%%%%%%%%%%%%%%%%%%%%%%%%%%%%%%%%%%%%%%%%%%%%%%%%%%

\subsection{Non-contractible diagrams}

As advertised in the beginning of this section, we will now show that
non-contractible diagrams, figs. 13 and 14, yield a finite
contribution to torus anomalous dimension. As one can observe in
fig. 14,  to join the legs of the interaction vertex while both interaction
loops surround the same cycle of the torus, it is necessary that one
of the 4 possible $Z$-blocks be absent (block 3 is absent in fig. 14). 
Therefore the
multiplicity of this class is $1/J$ lower than the semi-contractible
class, that is $\O(J^3)$. However phase summation will not change this
order essentially because upper and lower legs of the interaction
vertex are separated by a macroscopic number of $Z$ lines. One
concludes that non-contractible diagrams are also $\O(J^3)$ \ie
leading order. 

%%%%%%%%%%%%%%%%%%%%%%%%%%%%%%%%%%%%%%%%%%%%%%%%%%%%%%%%%%%%%%%%%%%%%%%%%%%%%%%
\begin{figure}[htb]
\centerline{ \epsfysize=6cm\epsffile{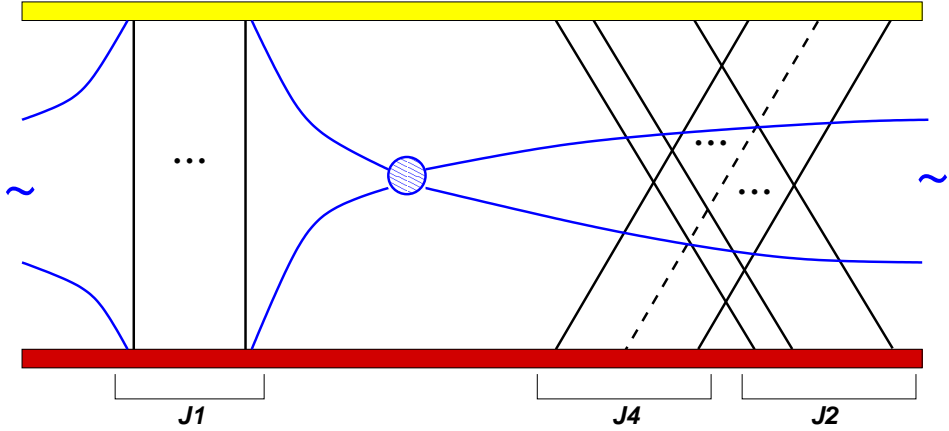}}
\caption{Non-contractible diagrams on the cylinder.}
\label{FIG13}
\end{figure}
%%%%%%%%%%%%%%%%%%%%%%%%%%%%%%%%%%%%%%%%%%%%%%%%%%%%%%%%%%%%%%%%%%%%%%%%%%%%%%

fig. 15 shows non-contractible external gluon diagrams.  
Having gained experience with previous calculations one can
immediately write down the contribution of the total D-vertex and
external gluons ( for the fixed position of $\phi$ shown in figures
) as, 
%%%%%
$$ \quart (1-q^{J_1})(1-\bar{q}^{J_1})
\bar{q}^{J_2}\<\Tr(U_{\m}(x)\bar{U}_{\n}(y))\> G(x,y)^J N^{J-3}.$$
%%%%%
This result should be summed over all positions of the scalar impurity
$\phi$ and finally over $J_1,\dots J_4$. Clearly no relative phase will
be associated when $\phi$ is in the first vertical block in
fig. 13. When it is in the second block, relative distance of $\phi$
and $\bar{\phi}$ to the interaction vertex is $J_3$, hence a nontrivial
phase, $q^{J_3}$ arises. The last case, when $\phi$ propagator is in
the third block was already considered above and yields the phase
$q^{-J_2}$. Replacing the sum over $J_i$ (with  $J_1+J_2+J_3=J$)
by an integral over $j_i = J_i/J$ (with $j_1+j_2+j_3=1$),
one arrives at the integral
%%%%%%%%%%%%%%%%%%%%
\be\la{jint2}
\int_0^1 \ud j_1
dj_2 dj_3 \,\delta(j_1+j_2+j_3-1)   
(j_2e^{2\pi i n j_3} + j_3 e^{-2\pi i n j_2} + j_1)
|1-e^{2\pi i n j_1}|^2
= \frac 13+ \frac{5}{2\pi^2 n^2}
\ee
%%%%%%%%%%%%%%%%%%%%
(for $n\neq0$). Using the result for current correlator from
Appendix A, one finds the following D-term and external gluon
contribution from the non-contractible diagrams, 
%%%%%%%
\be\la{torus2}
\<O^n_{\m}(x)\bar{O}^m_{\n}(y)\> \to
 (\frac 13+ \frac{5}{2\pi^2 n^2})G(x,y)^J
\ln (\Lambda^2(x-y)^2)\frac{J_{\m\n}(x-y)}{(x-y)^2}.
\ee      
%%%%%%%

%%%%%%%%%%%%%%%%%%%%%%%%%%%%%%%%%%%%%%%%%%%%%%%%%%%%%%%%%%%%%%%%%%%%%%%%%%%%%%%
\begin{figure}[htb]
\centerline{ \epsfysize=6cm\epsffile{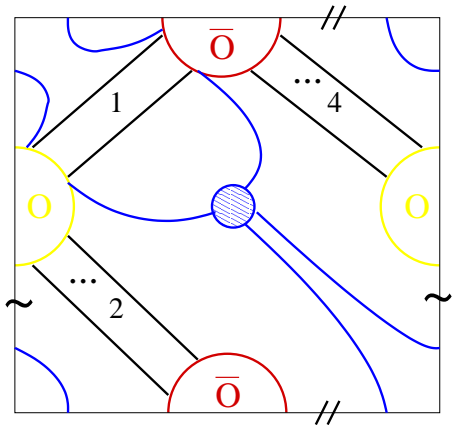}}
\caption{A Non-contractible diagram on the periodic square. 
Note that 3rd block of $Z$-lines is missing.}
\label{FIG14}
\end{figure}
%%%%%%%%%%%%%%%%%%%%%%%%%%%%%%%%%%%%%%%%%%%%%%%%%%%%%%%%%%%%%%%%%%%%%%%%%%%%%%

The non-contractible F-term contribution in fig. 13 arises when
$\phi$ (and $\bar{\phi}$ ) impurity is at the first or last positions
of the first block where one replaces the total vertex with an F-term
quartic vertex. Note that the integral over this vertex gives the logarithmic
scaling, 
%%%%%%%
\be\la{lambdaint}
\frac{1}{(4\pi^2)^4}
\int \frac{\ud^4 u}{(x-u)^4(y-u)^4}=2\pi^2 \ln (\Lambda^2(x-y)^2)
G(x,y)^2.
\ee
%%%%%%%
Now, one should dress this diagram by all possible locations of the
derivatives. When both $\p_{\m}$ and $\p_{\n}$ hit the same line, the
phase summation is equivalent to the situation discussed above. A
double derivative line replaces the $\phi$ impurity whose position is
to be summed over as in (\ref{jint2}) and one again finds out the
factor $(\frac 13+ \frac{5}{2\pi^2 n^2})$ together with the space-time dependence, 
$\p_{\m}\p_{\n}\frac{1}{(x-y)^2}$. The case where    
$\p_{\m}$ and $\p_{\n}$ hits different lines is handled in the same
way as in section 2. One first considers 
a fixed position of $\p_{\m}$, say $l$, and sum over position of $\p_{\n}$ from
$l'=0$ to $J+1$ with the condition $l'\ne \sigma(l)$. This yields a
factor $-q^l\bar{q}^{\sigma(l)}$ which is then summed over $l$ and
finally over $J_1,\dots J_4$ resulting in the same phase factor
(\ref{jint2}) up to a minus sign but with a different space-time dependence, 
$\p_{\m}\frac{1}{(x-y)^2}\p_{\n}\frac{1}{(x-y)^2}$. Combining these
cases one gets, 
%%%%%%%
$$
(\frac 13+ \frac{5}{2\pi^2 n^2})\frac{J_{\m\n}(x-y)}{(x-y)^2}
$$      
%%%%%%%
where we used (\ref{Jmunu2}). Therefore F-term contribution to anomalous part
of the torus correlator is exactly the same as (\ref{torus2}) and the
total result involving D-term, external gluon and F-term contribution 
simply becomes, 
%%%%%%%%%%%%%%%%%%%%%%%%%%%%%%%%%%%%%%%%%%%%%%%%%%%%%%%%%%%%%%%%%%%%%%%%
\be\la{torus3}
\<O^n_{\m}(x)\bar{O}^m_{\n}(y)\>_{D-term} \to
 \frac{\lambda'g_2^2}{4\pi^2}(\frac 13+ \frac{5}{2\pi^2 n^2})G(x,y)^J
\ln (\Lambda^2(x-y)^2)\frac{2J_{\m\n}(x-y)}{(x-y)^2}.
\ee
%%%%%%%%%%%%%%%%%%%%%%%%%%%%%%%%%%%%%%%%%%%%%%%%%%%%%%%%%%%%%%%%%%%%%%%%
This torus dimension is exactly the same as torus anomalous dimension
of BMN operators, \cite{Constable}.   

%%%%%%%%%%%%%%%%%%%%%%%%%%%%%%%%%%%%%%%%%%%%%%%%%%%%%%%%%%%%%%%%%%%%%%%%%%%%%%
\begin{figure}[htb]
\centerline{ \epsfysize=4cm\epsffile{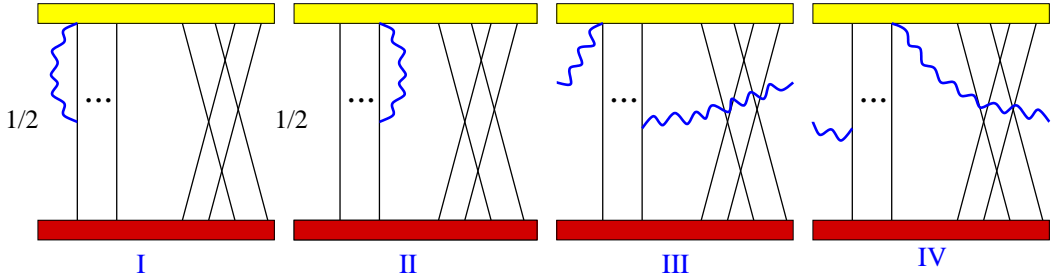}}
\caption{External gluon interactions with non-contractible topology.}
\label{FIG15}
\end{figure}
%%%%%%%%%%%%%%%%%%%%%%%%%%%%%%%%%%%%%%%%%%%%%%%%%%%%%%%%%%%%%%%%%%%%%%%%%%%%%%

\subsection{Special diagrams}

All of the topological classes of Feynman graphs that have been discussed so 
far were available both for F-term and D-term  parts of the
Lagrangian (\ref{lagr}). However, the special Feynman graphs on the torus are
formed when the interaction loops wind around different cycles and are
present only if the interaction is a D-term quartic vertex (and their
external gluon cousins). To see this, one should specify the
orientation of the scalar propagator line $Z\bar{Z}$ (and
$\phi\bar{\phi}$) by putting an arrow on it (not to be confused
by derivatives). We choose the convention where scalar propagation
is from $O$ towards $\bar{O}$. With specification of
the orientations, the F-term
and D-term quartic vertices can be represented as in fig. 16. 
One observes that the vertex where {\emph{ adjacent lines have the 
opposite orientation}} is only possible for D-terms. 
Using such a vertex one can draw 4 different special graphs on a
torus. One of these possibilities is shown in figs. 17 or 18. 
Here the shaded circle represents the
total vertex, fig. 2 as before. Special graphs can also be formed by
external gluons as in fig. 19. In general, special graphs are formed by 
combining either first or last lines of blocks 1 and 3 or blocks 2 and 4.  
%%%%%%%%%%%%%%%%%%%%%%%%%%%%%%%%%%%%%%%%%%%%%%%%%%%%%%%%%%%%%%%%%%%%%%%%%%%%%%
\begin{figure}[htb]
\centerline{ \epsfysize=4cm\epsffile{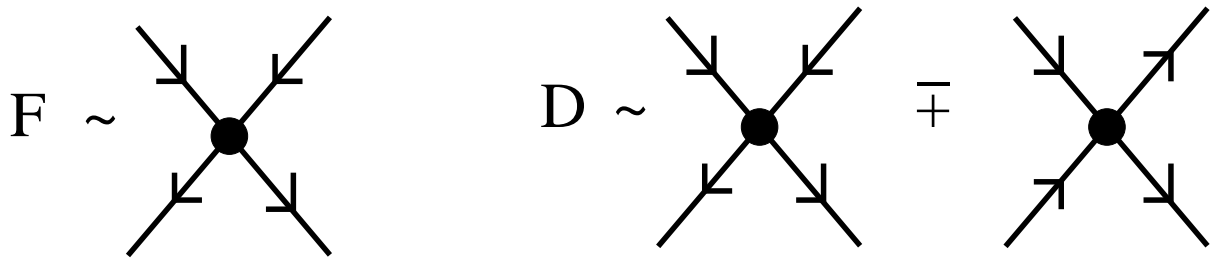}}
\caption{Orientations of F-term and D-term quartic vertices.}
\label{FIG16}
\end{figure}
%%%%%%%%%%%%%%%%%%%%%%%%%%%%%%%%%%%%%%%%%%%%%%%%%%%%%%%%%%%%%%%%%%%%%%%%%%%%%%

However, one makes a disturbing observation about special graphs:
They are $\O(J^4)$ therefore all of the graphs we have considered so
far are sub-leading with respect to them! Even worse, this extra power
of $J$ seems to be unsuppressed in the BMN limit, hence the presence
of such graphs imply the breakdown of BMN perturbation theory!?
Hopefully, as we shall demonstrate next, contribution of special graphs to
the anomalous dimension is zero when one adds up all such possible
graphs (fig. 17) just as in the semi-contractible case. 

%%%%%%%%%%%%%%%%%%%%%%%%%%%%%%%%%%%%%%%%%%%%%%%%%%%%%%%%%%%%%%%%%%%%%%%%%%%%%%
\begin{figure}[htb]
\centerline{ \epsfysize=6cm\epsffile{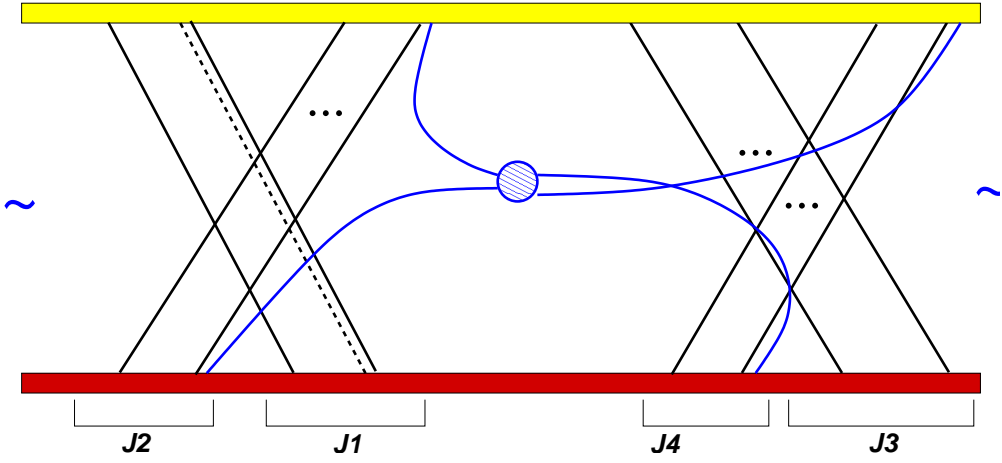}}
\caption{Special diagrams shown on a cylinder.}
\label{FIG17}
\end{figure}
%%%%%%%%%%%%%%%%%%%%%%%%%%%%%%%%%%%%%%%%%%%%%%%%%%%%%%%%%%%%%%%%%%%%%%%%%%%%%%

Let us consider a fixed position of $\phi$ at the last line of the
first block and fixed $J_1,\dots,J_4$. By use of trace identities
given in Appendix C and $q$-derivation tricks described above, 
one can easily boil down the special D-graphs into our
familiar $\<J\bar{J}\>$ correlator. Let us first consider the special contribution to 
the chiral primary correlator, $\O_{\phi}^J$. The trace 
identities show that fig. 17 
$$\sim (B+A)G^{J+2}N^{J+1}$$ hence special graph contribution to chiral
primary correlator vanishes by non-renormalization theorem. Reader
will find the details of this calculation in Appendix B. Next we put
in the $q$-derivatives on this graph to obtain the special contribution to 
$\<O_{\m}^n\ol{O}_{\n}^m\>$ and observe that the only positions
which yield a non-vanishing result is when both $\p_{\m}$ and
$\p_{\n}$ act on the total vertex. Proof of this fact is exactly analogous to 
our argument in section 4.2. 
The algebraic tricks familiar from previous calculations are then used
to express the result as 
$$\sim
q^{-J_2}\bar{r}^{J_1}(1-q^{-J_2-J_3})(1-\bar{r}^{-J_3-J_4})\<J_{\m}(x)\bar{J}_{\n}(y)\>G^{J}N^{J+1}.
$$ 
Similarly, the external gluon contributions shown in fig. 19 can be
shown to have the same form and total result---which follows from combining D-term, external gluon and 
self-energy contributions---becomes, 
$$\sim
q^{-J_2}\bar{r}^{J_1}(1-q^{-J_2-J_3})(1-\bar{r}^{-J_3-J_4})\<U_{\m}(x)\bar{U}_{\n}(y)\>G^{J}N^{J+1}.
$$ 

%%%%%%%%%%%%%%%%%%%%%%%%%%%%%%%%%%%%%%%%%%%%%%%%%%%%%%%%%%%%%%%%%%%%%%%%%%%%%%
\begin{figure}[htb]
\centerline{ \epsfysize=6cm\epsffile{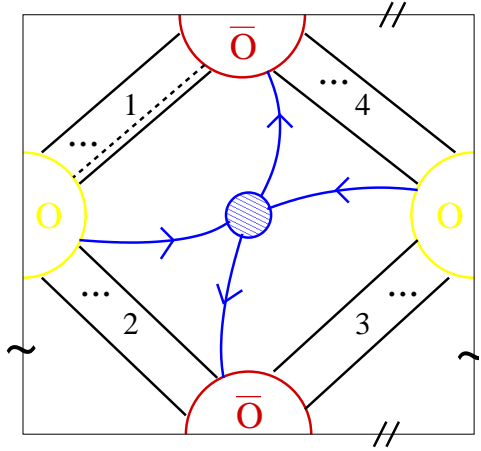}}
\caption{Periodic square representation of Special diagrams.}
\label{FIG18}
\end{figure}
%%%%%%%%%%%%%%%%%%%%%%%%%%%%%%%%%%%%%%%%%%%%%%%%%%%%%%%%%%%%%%%%%%%%%%%%%%%%%%

This was for the diagram in fig. 17. A second special graph is obtained when the legs of 
the total vertex stretchs out into the last line of block 1 and last of block 3. Similarly
a third graph is formed by first line in block 2 with first of 4 and a fourth graph by the 
first of 1 with first of 3.    
Let us now read off the phase factors of these four graphs respectively, 
%%%%%
\bea &(1-q^{-J_4-J_3})(1-\bar{r}^{-J_1-J_4})q^{-J_2}\bar{r}^{J_1},\;\;\;
&(1-q^{-J_2-J_3})(1-\bar{r}^{-J_3-J_4}),\nonumber\\
&(1-q^{-J_2-J_3})(1-\bar{r}^{-J_1-J_2})\bar{r}^{-J_3-J_4},\;\;\;
&(1-q^{-J_2-J_1})(1-\bar{r}^{-J_1-J_4})q^{J_1}\bar{r}^{J_1},\nonumber
\eea
%%%%
respectively. Hence the contribution from 1st graph cancels out 4th graph 
and the 2nd cancels out th 3rd. {\emph{We conclude that contribution of special
diagrams to both the vector anomalous dimension (the case $n=m$) and
the operator mixing ( the case $n\ne m$) vanishes}} although it seems
to be divergent as $J\to\infty$ at first sight. 
This shows that the only non-vanishing contribution
is arising from non-contractible class of diagrams and the total
correlator including $\O(\l')$ corrections both at the planar and the
torus levels can now be written as, 
%%%%%%%%%%%%%%%%%%%%%%%%%%%%%%%%%%%%%%%%%%%%%%%%%%%%%%%%%%%%%%%%%%%%%%
\bea\la{TOTAL}
\<O^n_{\m}(x)\bar{O}^n_{\n}(y)\> &=& 
 \left\{(1+g_2^2A_{nn})(1-n^2\lambda'\ln(\Lambda^2(x-y)^2)
+\frac{\lambda'g_2^2}{4\pi^2}
\left(\frac 13+ \frac{5}{2\pi^2 n^2}\right)
\ln(\Lambda^2(x-y)^2)\right\}\nonumber\\
{}&&\qquad \times G(x,y)^{J+2}\frac{2J_{\m\n}(x,y)}{(x-y)^2}.
\eea
%%%%%%%%%%%%%%%%%%%%%%%%%%%%%%%%%%%%%%%%%%%%%%%%%%%%%%%%%%%%%%%%%%%%%%

This result clearly shows that the total contribution to the anomalous
dimension of the vector type operator at $\O(g_{YM}^2)$ and up to
genus-2 level is exactly the same as the BMN anomalous dimension, that
is, 
%%%%%%%%%%%%%%%%%%%%%%%%%%%%%%%%%%
\be\la{totandim}
\Delta =
J + 2 + \lambda'n^2
-\frac{g_2^2\lambda'}{4\pi^2}
\left(
\frac 13+ \frac{5}{2\pi^2 n^2}
\right).
\ee
%%%%%%%%%%%%%%%%%%%%%%%%%%%%%%%%%%

As mentioned before, from the string theory point of view, the torus anomalous dimension is
identified with the genus-one mass renormalization of the
corresponding state. 

%%%%%%%%%%%%%%%%%%%%%%%%%%%%%%%%%%%%%%%%%%%%%%%%%%%%%%%%%%%%%%%%%%%%%%%%%%%%%%
\begin{figure}[htb]
\centerline{ \epsfysize=4cm\epsffile{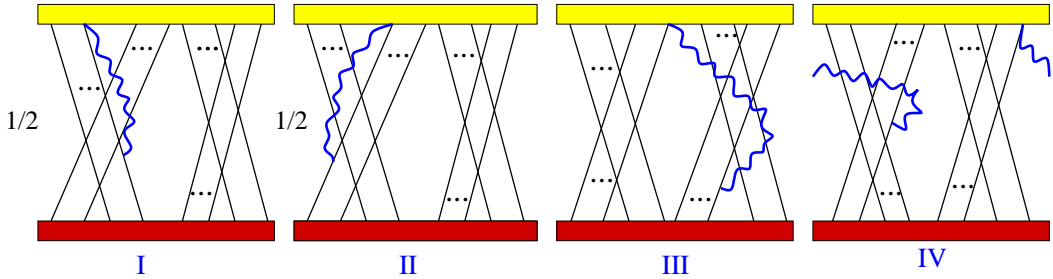}}
\caption{External gluons with special diagram topology.}
\label{FIG19}
\end{figure}
%%%%%%%%%%%%%%%%%%%%%%%%%%%%%%%%%%%%%%%%%%%%%%%%%%%%%%%%%%%%%%%%%%%%%%%%%%%%%%

Let us briefly describe how the non-vanishing of torus level anomalous 
dimension implies the non-vanishing of the $\O(\l')$ interacting three-point 
functions, 
%%%%
$$\<\bar{O}_{\m}^{n,\,J} O_{\n}^{m,\,J'} O^{J-J'}\> \qquad {\textrm{and}} \qquad 
 \<\bar{O}_{\m}^{n,\,J} O_{\n}^{J'} O_{\phi}^{J-J'}\>$$
through the unitarity sum. Here the relevant supergravity operators are defined in 
(\ref{sgp}) and (\ref{omu}). As mentioned in the introduction, 
this is puzzling since string field theory result of \cite{MN2} shows 
that RHS of (\ref{7conj}) vanishes 
for vector operators. 

It was argued in \cite{Constable} that one can handle the 
string interactions effectively with non-degenerate perturbation theory 
of a quantum mechanical system. Unitarity sum gives the following 2nd order shift 
in the energy of the string state with momentum $n$, 
%%%%
\be\la{enshift}
E_n^{(2)} = \sum_{m\neq n}
\frac{|\langle i'| P^- |j'k' \rangle|^2}{E_n^{(0)}-E_m^{(0)}}.
\ee     
%%%%
Here $|i'\>$ is the string excitation with momentum $n$ and $|j'k'\>$ represents 
all possible intermediate states with momentum $m$. In the case of 
$|i'\> = \a_n^{\phi\,\dg}\a_{-n}^{\m\,\dg}|0,p^+\>$ which is the dual of 
vector operator (\ref{vectorop1}), there are two possibilities for the 
intermediate states:  
\begin{enumerate}
\item
$|j'\> = \a_m^{\phi\,\dg}\a_{-m}^{\m\,\dg}|0,p_1^+\>$ and 
$|k'\> = |0,p_2^+\>$ with $p_1^++p_2^+=p^+$. 
Corresponding operators are, $O_{\n}^{m,\,J'}$ and $O^{J-J'}$ respectively. 
\item
$|j'\> = \a_0^{\phi\,\dg}|0,p_1^+\>$ and 
$|k'\> = \a_0^{\m\,\dg}|0,p_2^+\>$. Corresponding operators are the BPS operators, 
$O_{\n}^{J'}$ and $O_{\phi}^{J-J'}$. 
\end{enumerate} 
Therefore the sum in (\ref{enshift}) involves a sum over these two cases together 
with sub-summations over $m$ and $J'$. Vanishing of $\langle i'| P^- |j'k' \rangle$ 
for both of the cases above implies that 
{\emph{$E_n^{(2)}=(\Delta-J)_{torus}=0$ for the vector operator}}. A loophole in this
argument is that we only considered the cubic string vertex in the effective description 
whereas the {\emph{contact terms}} may also contribute the mass renormalization of the 
string states hence give rise to a non-zero 
torus level anomalous dimension in the dual theory. 
We come back to this issue in the last section.   

\section{A SUSY argument}

The fact that BMN and vector operators (which belong to separate $SO(4)$ sectors 
of the gauge theory) have equal anomalous dimensions 
both at planar and torus levels suggests that there might be a ${\cal N}=4$ 
SUSY transformation relating these two operators. Whereas the equality of 
the planar anomalous dimensions of these operators is required by the
consistency of BMN conjecture, there is no {\it a priori} reason to
believe that this equality persists at higher genera. A SUSY map, however, 
would protect 
$\Delta_{BMN}-\Delta_{vector}=-1$ at all loop orders and all genera. 

In this section, we will see that indeed there is such a transformation 
which maps the BMN operator onto vector operator plus a correction term. 
We will argue that the correction is negligible in the BMN limit and hence 
expect the equality of anomalous dimensions, {\emph{both at planar and torus levels}}.  

The supersymmtery transformations of ${\cal N}=4$ SYM
has recently been derived in \cite{Okuyama}. In $SU(4)$ symmetric notation, 
the transformations of the scalars and chiral spinors read,
%%%%%%%%%
\bea
\d_\e X^{AB}&=&-i(-\bar{\e}_-^A\q_+^B+\bar{\e}_-^B\q_+^A
+\e^{ABCD}{\q}_{+C}\e_{-D})\la{susyX}\\
\d_\e\q_+^A&=&\half F_{\mu\nu}\gamma^{\mu\nu}\e_+^A
+2D_{\mu}X^{AB}\gamma^{\mu}\e_{-B}
+2i[X^{AC},X_{CB}]\e_+^B\la{susyl}.
\eea
%%%%%%%%%
in which $A=1,\dots 4$ is an $SU(4)$ index and $X^{AB}=-X^{BA}$.

We will use these transformation rules in a somewhat schematic way, since the information 
we need can be obtained more simply by classifying all fields and supercharges with 
respect to the decomposition  $SU(4)\to U(1)\times U(1)\times U(1)$. The three commuting 
$U(1)$ charges can be viewed as $J_{12}$, $J_{34}$ and $J_{56}$ in $SO(6)$. 
All fermionic quantities are taken as 2 component Weyl spinors. The four spinor fields 
are denoted by $\q_{\phi},\,\q_{\psi},\,\q_{Z},\,\q_{A_{\m}}=\l$ where the subscript indicates 
their bosonic partner in an ${\cal N}=1$ decomposition of ${\cal N}=4$ SUSY. The fermionic 
transformation rule above may be interpreted as, 
$$\{\ol{Q}_{+B},\q^A\}\sim F_{\m\n}\gamma^{\m\n}\delta_A^B+\cdots,$$ 
showing that $\ol{Q}_{+A}$ has the same $U(1)$ quantum numbers as $\q^A$. In general fermions 
and anti-fermions have opposite $U(1)$ charges, as in the case of conjugate bosons. 
The product of these 3 charges is positive on the $\ol{Q}_{+A}$. With these remarks in view, 
we can write the following table of $U(1)$ charges. 
\newline

%%%%%%%%%%%%%%%%%%%%%%%%%
\begin{tabular}{|c||c||c||c||c||c||c||c||c||c||c||c||c|}
\hline
{}&$\phi$&$\psi$&$Z$&$A_{\m}$&$\q_{\phi}$&$\q_{\psi}$&$\q_{Z}$&$\l$&$Q^1$&$Q^2$&$Q^3$&$Q^4$\\
\hline
$J_{12}$&1&0&0&0&1/2&-1/2&-1/2&1/2&-1/2&1/2&1/2&-1/2\\\hline
$J_{34}$&0&1&0&0&-1/2&1/2&-1/2&1/2&1/2&-1/2&1/2&-1/2\\
\hline
$J_{56}$&0&0&1&0&-1/2&-1/2&1/2&1/2&1/2&1/2&-1/2&-1/2\\
\hline
\end{tabular}
%%%%%%%%%%%%%%%%%%%%%%%%%
\newline

We now apply the transformation rules (\ref{susyX}) and (\ref{susyl}) in the $U(1)\times U(1)\times U(1)$ 
basis in which all transformations which conserve the $U(1)$ charges are allowed. 
Consider the action of $Q^2_{\a}$ on the BMN operator (\ref{BMNop}).  
We see that $\phi$ and $Z$'s are left unchanged 
whereas $\psi$ is transformed into a gaugino $\l$, \ie
%%%%
$$ [Q^2, O_{\phi\psi}^n] \propto
\sum_{l=0}^J e^{\frac{2\pi i nl}{J}}\Tr(\phi Z^l \l Z^{J-l})$$
%%%%
Next we act on this with another anti-chiral supercharge $\ol{Q}^3_{\a}$ with quantum numbers
$(-1/2, -1/2, +1/2)$. According to table 1 and transformation rules given in (\ref{susyl}),  
$Z$'s again remain unchanged, $\l$ is transformed into 
$D_{\m}Z$\footnote{The last term in (\ref{susyl}) which is quadratic in the scalars does not 
give correct quantum numbers for $J_{12}\dots J_{56}$ hence is not present.}
and $\phi$ is transformed into $\bar{\q}_{\psi}$, i.e.
%%%%
\bea
\{\ol{Q}^3,[Q^2, O_{\phi\psi}^n]\} &\propto&
\sum_{l=0}^J e^{\frac{2\pi i nl}{J}}\Tr(\phi Z^l (D_{\m}Z) Z^{J-l})  \la{redmix}\\
{}&&+\sum_{l=0}^J e^{\frac{2\pi i nl}{J}}\Tr(\bar{\q}_{\psi} Z^l \l Z^{J-l})
\equiv \widetilde{O}_{\m}^n+O_f^n.\nonumber
\eea
%%%%
Therefore supersymmetry guarantees that $\widetilde{O}_{\m}^n+O_f^n$ has the same $\Delta-J$ with the BMN 
operator. 

Note that the first term is not quite the same as the vector operator of (\ref{vectorop1}) 
but there are two differences. However, for large but finite $J$, the difference between contributions of 
$\widetilde{O}_{\m}^n$ and $O_{\m}^n$ to the correlators is $\O(1/J)$. 
This is because the exceptional piece in (\ref{vectorop1}) 
where $D_{\m}$ is acting on the impurity $\phi$ is $\O(1/J)$ with respect to the first term of (\ref{vectorop1}) hence
negligible in the dilute approximation. Secondly, the difference between the definitions of $q$ for $\widetilde{O}_{\m}^n$
and  $O_{\m}^n$, \ie $q^J=1$ and $q^{J+2}=1$ respectively, is also $\O(1/J)$.

Now consider computing the dimension of $\widetilde{O}_{\m}^n+O_f^n$ 
at the planar level. This would be the same as the dimension of only $\widetilde{O}_{\m}^n$ provided that 
the transition amplitude $\<\widetilde{O}^n_{\m}\bar{O}_f^m\>$ is negligible. Let us first 
consider the correlator $\<O_{\m}^n\bar{O}_f^m\>$ instead. 
Above we explained that the difference between the contributions of 
$\widetilde{O}^n_{\m}$ and $O^n_{\m}$ to the anomalous dimension is $\O(1/J)$, therefore 
conclusions made for $\<O_{\m}^n\bar{O}_f^m\>$ will also be valid for $\<\widetilde{O}_{\m}^n\bar{O}_f^m\>$ 
as $J\to\infty$ in the BMN limit. The leading contribution to 
this transition amplitude in $\O(\l')$ arises from $Z$-$\l$-$\bar{\q}$ (5th term in (\ref{lagr})) and the Yukawa
interaction (6th term in (\ref{lagr})): One of the $Z$'s in $O_{\m}^n$ 
splits into a $\q_Z$ and $\bar{\l}$ and $\q_Z$ gets absorbed by $\phi$ turning into 
$\bar{\q}_{\psi}$ through the Yukawa interaction. See fig.20 for the analogous interaction at the torus level.   
Note that contribution to the {\emph{planar}} dimension
requires the $Z$ that is taking place in the interaction and $\phi$ be adjacent.  
Note also that the derivative in $O_{\m}^n$ can be at any position. 
Before acting with the $q$-derivative, 
the integration over the internal vertices together with the scalar propagators yields, 
%%%%
$$\sim\ln((x-y)^2\Lambda^2)G^J.$$ 
%%%%
Since the anomalous dimension is the coefficient of the $\log$ term, anomalous contributions arise when 
the $\p_{\m}$ act on $G$'s but not on the $\log$. Therefore any position of the derivative in $O_{\m}^n$ 
in the planar diagram (also any position of the derivative in fig. 20 in torus case) 
gives the same contribution, 
%%%%
$$\sim\ln((x-y)^2\Lambda^2)G^{J+1},$$ 
%%%%
regardless it is acting on the fields participating in the interaction or not. Then 
the phase sum over the position of $\p_{\m}$ gives
(using $q^{J+2}=1$),
%%%%
$$\sum_{l=0}^{J+1} q^l = 0.$$
%%%%
Therefore the transition amplitude $\<O_{\m}^n\bar{O}_f^m\>_{planar}$ vanishes identically! 
As we described above this implies that, for large but finite $J$, 
$\<\widetilde{O}_{\m}^n\bar{O}_f^m\>_{planar}$ do not vanish but supressed with a factor of 
$1/J$ with respect to $\<\widetilde{O}_{\m}^n\bar{\widetilde{O_{\m}^n}}\>_{planar}$ in the BMN 
limit. Therefore we see that {\emph{supersymmetry together 
with large $J$ suppression is capable to explain why vector and BMN anomalous dimensions 
are equal at the planar level}}. 

%%%%%%%%%%%%%%%%%%%%%%%%%%%%%%%%%%%%%%%%%%%%%%%%%%%%%%%%%%%%%%%%%%%%%%%%%%%%%%
\begin{figure}[htb]
\centerline{ \epsfysize=6cm\epsffile{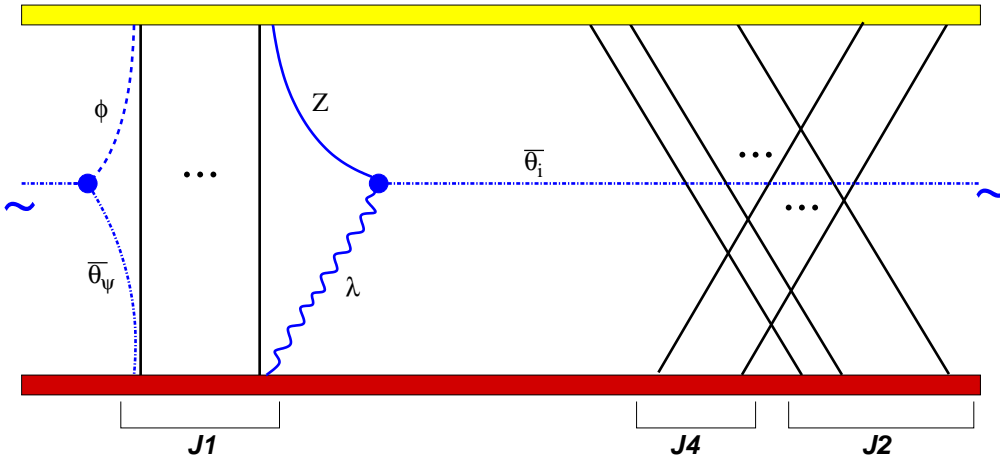}}
\caption{Torus level non-contractible contribution to $\<O^n_{\m}\bar{O}_f^m\>$ transition amplitude. 
Derivative in $O_{\m}^n$ can be placed on any line connected to $O_{\m}^n$, 
although it is not shown explicitly. There is a similar graph obtained by interchanging internal vertices.}
\label{FIG20}
\end{figure}
%%%%%%%%%%%%%%%%%%%%%%%%%%%%%%%%%%%%%%%%%%%%%%%%%%%%%%%%%%%%%%%%%%%%%%%%%%%%%%

This argument can easily be extended to the torus level. In section 5.4 it was proven
that the only torus level contribution for the vector correlator $\<O_{\m}^n\bar{O}_{\n}^m\>_{torus}$ comes 
from the non-contractible diagram, fig. 13. Recall that this diagram is of $\O(J^3)$ because 
there are three blocks of $Z$ lines 
and no phase suppression (unlike contractible or semi-contractible cases). Including the $1/(JN^2)$ 
normalization factor we found out that the diagram is of $\O(g_2^2)$. 
We want to see how the transition amplitude at the torus level, $\<\widetilde{O}_{\m}^n\bar{O}_f^m\>_{torus}$ goes
with $J$. Instead, let us again consider the correlator $\<O_{\m}^n\bar{O}_f^m\>_{torus}$. 
One can easily see that (with the same argument presented in the beginning of section 5) 
all possible torus diagrams of $\<O_{\m}^n\bar{O}_f^m\>$ can be divided into 
four seperate classes of section 5. Let us consider the non-contractible diagram for example. 
The diagram is shown in fig. 20. The derivative in $O_{\m}^n$ can be at any position. 
Hence the phase summation over the position of the derivative vanishes identically 
just as in the planar case. For the same reason the external gluon contribution vanishes as well 
(together with other torus diagrams: contractible, semi-contractible and special). 
One concludes that for large but finite $J$, 
$\<\widetilde{O}_{\m}^n\bar{O}_f^m\>_{torus}$ is again $1/J$ suppressed with respect to $\<O_{\m}^n\bar{O}_{\n}^m\>_{torus}$.
We see that {\emph{supersymmetry in the 
BMN limit, is also capable to explain the equality of vector and BMN anomalous dimensions 
at the torus level}}. However we emphasize that this equality is not exact but only holds 
in large $J$ limit. Therefore, whether this reasoning can 
be extended to higher orders in genus remains as an interesting question.

As an aside we state another important conclusion. The fact that the transition amplitude is negligible 
with respect to $\<O_{\m}^n\bar{O}_{\n}^m\>$ shows that {\emph{the vector operator $O_{\m}^n$ and the fermionic impurity 
BMN operator $O_f^n$ has the same planar and torus anomalous dimensions}}. This gives an easy method to generate 
all BMN operators which carry the same anomalous dimension as the scalar operator by acting on it with the supercharges 
$Q^I$ arbitrary times {\emph{and}} making sure that 
the transition amplitudes among all of the pieces in the end-product is negligible in the BMN limit.

\section{Discussion and outlook}  

In this paper we computed the two-point function of vector impurity type BMN operator at planar and 
torus levels, for small $\l'$ (large $\mu p^+\alpha'$). In this regime, SYM is weakly coupled 
and we only considered interactions at $\O(\l')$ order in SYM interactions. Our result for the total 
anomalous dimension is given in eq. (\ref{totandim}). This turns out to be exactly the same as 
scalar impurity type anomalous dimension which was computed in \cite{Constable} 
{\emph{both at planar and torus levels}}. This result provides
two tests on the recent conjectures. This equality at the planar level constitutes a non-trivial 
check on the BMN conjecture. Secondly, the non-zero torus anomalous dimension is a field theory prediction 
which should match the string theory result for the mass renormalization of the vector states. 
We mentioned at the end of section 5 however that this non-zero torus dimension raises a puzzle since 
the string field theory cubic vertex for vector the string states vanishes\cite{MN2}. 
Our results are further supported by the SUSY argument given in the previous section. 

We would like to briefly address some possible resolutions of this contradiction between 
string field theory and gauge theory results. Generally speaking, 
there is another type of interaction in light-cone string field theory \cite{GSW2} 
apart from the cubic string vertex. This arises from the {\it{contact terms}} 
and was not taken into account in the calculation of \cite{MN2}. In the context of IIB strings 
in pp-wave background this issue was discussed in \cite{BN}. Contact terms arise from a quartic string 
vertex whose presence is required by supersymmetry \cite{Greensite}\cite{Green}. Contribution of
contact terms to mass renormalization is $\O({g'_s}^2)$ and there seems no {\it a priori} 
reason to ignore it. In case these terms are indeed non-negligible they might give rise to a non-zero 
torus anomalous dimension in the dual field theory.  

Another resolution \footnote{I am Grateful to L. Motl for mentioning this idea to me.} 
of the gauge/string contradiction would be that peturbative 
gauge theory calculations for the interacting three-point function are not capable to probe 
the short distance ($< 1/\mu$) physics on the world-sheet. 
Recall that \cite{MN2} it is the {\it{prefactor}} of cubic vertex which suggests the vanishing of 
$\<i'|P^-|j'\>|k'\>$ in case of the vector state: Spradlin and Volovich have pointed out that 
the short distance limit on the world-sheet and the weak gauge coupling limit ($\mu\to\infty$) 
do not commute. To be able to obtain the {\it{prefactor}} one should first take the short distance 
limit. Then one takes large $\mu$ limit to obtain an expression for the weakly coupled 
three-point function. This procedure expects vanishing of $\<i'|P^-|j'\>|k'\>$. 
On the other hand, exchanging the limits, hence loosing the contribution of {\it{prefactor}} would 
suggest non-zero interacting three-point function also for the vector operator. It is a possibility 
that perturbative SYM is not able to ``discover'' the {\it{prefactor}} of string field theory but able to 
see only a $1/\mu$ expansion of the delta-functional. This would be another line of reasoning to explain 
why our perturbative calculation produced a non-zero torus anomalous dimension for the vector operator. 
Clearly, a perfect understanding of the map between weakly coupled string/gauge theories should 
resolve this apparent contradiction.  

A number of directions for further study of these issues are the following. 
One can take a direct approach and compute the contact terms in string field theory to compare its 
contribution to torus level mass renormalization with the contribution of non-contractible diagrams 
to the torus anomalous dimension. A similar strategy in gauge theory side is to obtain the interacting 
three-point function of the vector operators as a direct test of (\ref{7conj}). It is also desirable to go 
beyond $\O(\l')$ and obtain a non-perturbative formula for the anomalous dimension of the vector operators 
with a similar calculation as in \cite{BMN} for the case of scalar operators.  

~

\centerline{\bf Acknowledgements}

It is a pleasure to thank David Berenstein, Matt Headrick, Shiraz Minwalla, Juan Maldacena, 
Horatiu Nastase and Witek Skiba for useful discussions. I must specially thank to 
Dan Freedman and Lubos Motl for their early participation and constant help during the preperation 
of the paper. Section 5.6 is worked out together with them.  
This work is supported by funds provided by the D.O.E. 
under cooperative research agreement $\#$DF-FC02-94ER40818. 

\newpage
    
\appendix

\section{D-term and external gluon contributions at $\O(\l')$}

The aim of this appendix is to prove equation (\ref{andim1}) showing 
the planar level contribution to the vector anomalous dimension
which arise from the D-term part of the lagrangian, (\ref{lagr}). In
section 3 we explained how one can express this result in terms of
the the correlator of a non-conserved current, $J_{\m}=Z\ad Z$. Here,
we shall compute D-term contributions to this correlator at the one 
loop level. In the following we will first show that D-term quartic
vertex do not contribute to $\<J\bar{J}\>$ at all. Then we will
explain how to compute the self-energy contributions. Finally we shall
consider the contribution arising from the gluon exchange graph fig. 4I
by employing the trick of relating it to simpler diagrams  figs 4II, 4III, 
4IV as we described at the end of section 3.2. 

The most direct and painless way to compute these Feynman diagrams is
the beautiful method of differential renormalization (DR)
\cite{difrenbible}. The main idea is to compute $n$-point functions
$\<O_1(x_1)\cdots O_k(x_k)\>$ directly in space time rather than
Fourier transforming to momentum space, and adopting a certain
differential regularization scheme when the space-time expressions become
singular, \ie as $x_i\to x_j$. Note that, away from the contact points  
$x_i\to x_j$, the $n$-point function is well-defined and can be Fourier
transformed back to momentum space. However as $x_i$ approachs to
$x_j$ for $i\ne j$, most expressions become too singular to admit a
Fourier transform. Yet, one can easily rewrite the singular
expressions in terms of derivatives of less singular expressions hence
render the Fourier transform possible. The only such rewriting we will
use here is the following formula \cite{difrenbible}, 
%%%%
\be\la{dr1}
\frac{1}{(x_1-x_2)^2} = -\quart\Box\frac{\ln((x_1-x_2)^2\Lambda^2)}
{(x_1-x_2)^2}
\ee
%%%%
where $\Lambda^2$ is the renormalization scale. We also remind the
Green's equation in our conventions,  
%%%%
\be\la{dr2}
\Box\frac{1}{(x_1-x_2)^2}=-4\pi^2\delta(x_1-x_2).
\ee
%%%%
A nice feature of DR
is that one can adopt a renormalization scheme where one ignores all
tadpole diagrams, simply by setting them to zero. We will work with
the Euclidean signature throughout the appendices in which case the
space-time Feynman rules for ${\cal N}=4$ SYM are read off from
(\ref{lagr}). Scalar, gluon and fermion propagtors read 
%%%%
$$\frac{\delta_{ab}}{4\pi^2(x-y)^2};\;\;\;
\frac{\delta_{ab}\delta_{\m\n}}{4\pi^2(x-y)^2};\;\;\;
\frac{\delta_{ab}}{4\pi^2}\psh_x\frac{1}{(x-y)^2}.$$
%%%%
The interaction vertices are shown in fig. 21. 
%%%%%%%%%%%%%%%%%%%%%%%%%%%%%%%%%%%%%%%%%%%%%%%%%%%%%%%%%%%%%%%%%%%%%%%%%%%%%%
\begin{figure}[htb]
\centerline{ \epsfysize=6cm\epsffile{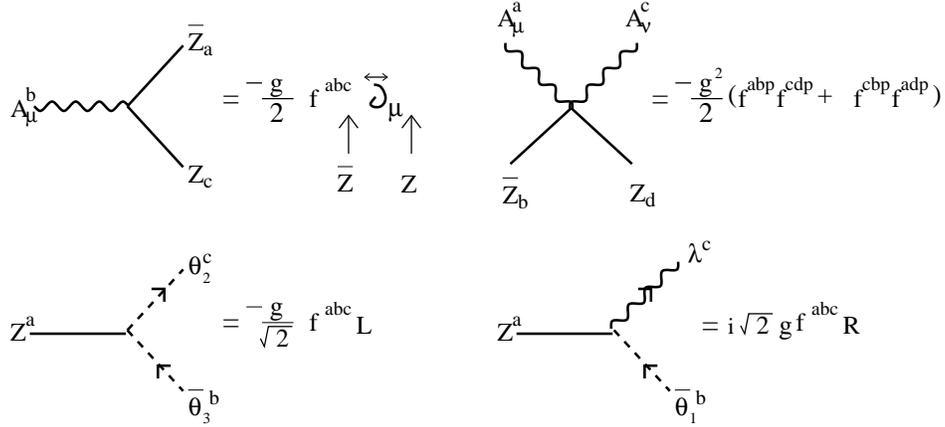}}
\caption{Feynman rules for vertices. Same rules hold when $Z$ 
is replaced by $\phi$. 
In the $\theta$-$Z$-$\bar{\theta}$ vertex, exchanging chiral 
fermion flavor 2 with 3 gives a minus sign and replacing $Z$ 
with $\bar{Z}$ changes the chirality projector from $L$ to $R$. 
The analogous $\bar{Z}$-$\bar{\lambda}$-$\theta$ vertex is 
obtained by replacing $R$ with $-L$.}
\label{figA1}
\end{figure}
%%%%%%%%%%%%%%%%%%%%%%%%%%%%%%%%%%%%%%%%%%%%%%%%%%%%%%%%%%%%%%%%%%%%%%%%%%%%%%

It is very easy to see that D-term contribution vanishes. 
%%%%%%%%%
\bea\la{Dterm}
\<\Tr(J_{\m}J_{\n})\>_D&=&\frac{g_{YM}^2}{4\pi^2}(f^{abd}f^{ace}+f^{abe}f^{acd})
\Tr(T^aT^bT^cT^d)\nonumber\\    
{}&&\int \ud^4u
(\frac{1}{(x-u)^2}\ad_{\m}\frac{1}{(y-u)^2}\ad_{\n}\frac{1}{(y-u)^2}=0.
\nonumber
\eea
%%%%%%%%%

To find the self-energy contribution to $\<J\bar{J}\>$ let us first
compute self-energy corrections to scalar propagator. These arise from
three sources: a gluon emission and reabsorbtion, 
chiral-chiral fermion loop and chiral-gluino fermion loop. 
We will not need the exact value of the first contribution as will be explained
below. Let us begin with chiral-criral loop which is shown in
fig. 22. Calling this graph $SE_1$, Feynman rules yield, 
%%%%
\bea
SE_1&=&-2\frac{1}{4\pi^2}f^{acd}(-\frac{g}{\sqrt{2}})f^{bdc}(\frac{g}{\sqrt{2}})
\int\ud^4u\ud^4v
\frac{1}{(x-u)^2}\frac{1}{(y-v)^2}\Tr[L\psh_u\frac{1}{(u-v)^2} 
R\psh_v\frac{1}{(u-v)^2}]\nonumber\\
{}&=&-2\frac{g_{YM}^2\delta^{ab}N}{4\pi^2}
\int\ud^4u\ud^4v
\frac{1}{(x-u)^2}\frac{1}{(y-v)^2}\psh^u_{\a}\frac{1}{(u-v)^2} 
\psh^v_{\a}\frac{1}{(u-v)^2}\nonumber
\eea
%%%%
where factor of 2 in the first line comes from summing over two
fermion flavors $\theta_1$ and $\theta_2$. We will show the evaluation
of integral here for future reference. By parts in $u$ gives,
%%%%
\bea\la{I1}
I_1(x,y)&=&\int\ud^4u\ud^4v\left\{\frac{1}{(x-u)^2}\frac{1}{(y-v)^2}(-4\pi^2)
\frac{\d(u-v)}{(u-v)^2}+\half\frac{1}{(x-u)^2}
\frac{1}{(y-v)^2}\Box\frac{1}{(u-v)^4}\right\}\nonumber\\
{}&\to&-\frac{1}{8}\int\ud^4u\ud^4v\frac{1}{(x-u)^2}\frac{1}{(y-v)^2}
\Box_u\Box_v\frac{\ln((u-v)^2\Lambda^2)}{(u-v)^2}\nonumber\\
{}&=&-\frac{1}{8}(-4\pi^2)^2\int\ud^4u\ud^4v\d(x-u)\d(y-v)
\frac{\ln((u-v)^2\Lambda^2)}{(u-v)^2}\nonumber\\
{}&=&-\frac{(4\pi^2)^2}{8}\frac{\ln((x-y)^2\Lambda^2)}{(x-y)^2}
\eea
%%%%
In passing to second line we omitted the first term which is supposed
to cancel out with tadpole contributions in DR. Second and third
equalities use (\ref{dr1}) and (\ref{dr2}). Hence,
%%%% 
\be\la{SE1}
SE_1=\frac{g_{YM}^2N}{4(4\pi^2)^2}\d_{ab}\frac{\ln((x-y)^2\Lambda^2)}{(x-y)^2}.
\ee
%%%%
An analogous computation for the gluino-chiral fermion loop gives, 
%%%% 
\be\la{SE2}
SE_2=-\frac{g_{YM}^2N}{2(4\pi^2)^2}\d_{ab}\frac{\ln((x-y)^2\Lambda^2)}{(x-y)^2}.
\ee
%%%%

Although we will not need it, let us give here the total self-energy
correction to the scalar propagator for reference
(including gluon emission-reabsorbtion),
%%%% 
\be\la{SE}
SE=-\frac{g_{YM}^2N}{8(4\pi^2)^2}\d_{ab}\frac{\ln((x-y)^2\Lambda^2)}{(x-y)^2}.
\ee
%%%%

%%%%%%%%%%%%%%%%%%%%%%%%%%%%%%%%%%%%%%%%%%%%%%%%%%%%%%%%%%%%%%%%%%%%%%%%%%%%%%
\begin{figure}[htb]
\centerline{ \epsfysize=4cm\epsffile{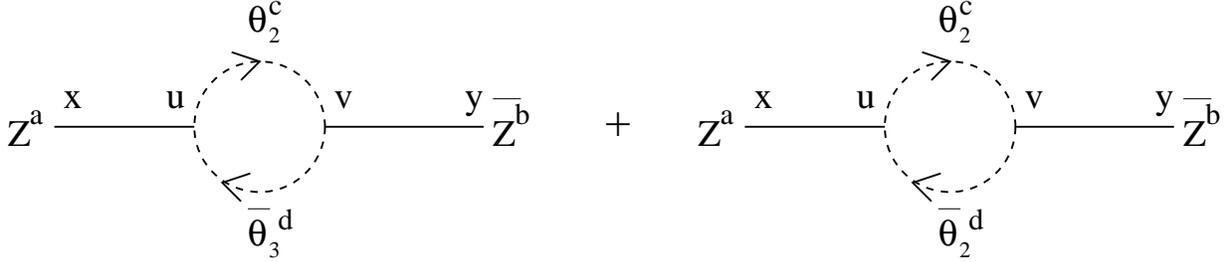}}
\caption{Chiral fermion loop contributions to self-energy of $Z$.}
\label{figA2}
\end{figure}
%%%%%%%%%%%%%%%%%%%%%%%%%%%%%%%%%%%%%%%%%%%%%%%%%%%%%%%%%%%%%%%%%%%%%%%%%%%%%%

Turning to gluon exchange contribution to $\<J\bar{J}\>$, we recall
our trick to express it as gluon exchange correction to gluon
propagator in scalar QED, fig. 4III,
%%%%
\bea\la{GE1}   
\<\Tr(J_{\m}(x)\bar{J}_{\n}(y))\>_{g.e.}&=&
\frac{\d_{ab}}{2}\<J^a_{\m}(x)\bar{J}^b_{\n}(y))\>_{g.e.}\nonumber\\
{}&=&-\frac{N^4}{g_{YM}^2}\<A_{\m}(x)A_{\n}(y)\>_{g.e.}
\eea
%%%%
where $J^a_{\m}=if^{abc}Z^b\p_{\m}Z^c$. In the second line we divided
out by a factor of $(g/2)^2$ to compensate for the coupling of
incoming and outgoing gluons to the loop and there is an overall $-1$
w.r.t. $\<J\bar{J}\>$ because of the antisymmetric derivative 
in scalar-gluon vertex. We also took into account the color factors at four vertices, 
$$f^{cge}f^{dac}f^{fdg}f^{ehf}=\delta^{ab}\frac{N^2}{2}.$$
Now, the sub-divergent piece of this diagram cancels
out the sub-divergent pieces of graphs I and II in fig 4. 
Hence the contribution to anomalous dimension is, 
%%%%
$$\<A_{\m}(x)A_{\n}(y)\>_{g.e.}\to 
-4\times {\mathrm{Fig.4I}}-2\times{\mathrm{Fig.4II}}.$$
%%%%
When we include th self-energy corrections to $\<J\bar{J}\>$ second
term will be canceled out
by gluon emission-reabsorbsion part of the self energies and one is
left with, 
%%%%
\be\la{totu}\<\Tr(J_{\m}\bar{J}_{\n})\>_{g.e.+s.e.}=-\frac{N^4}{g_{YM}^2}
\left\{-4\times{\mathrm{Fig.4a}}-2\times{F.S.E.}\right\}\ee 
%%%%
with ``FSE'' being only the fermion loop contributions to the self
energy of $\<J\bar{J}\>$,
%%%%
\be\la{FSE}
FSE=-\frac{g_{YM}^2N^4}{16\pi^2}\frac{J_{\m\n}}{(x-y)^2}\ln((x-y)^2\Lambda^2)G(x,y)^2.
\ee
%%%%
Note that we included $1/2$ factor coming from our counting of self
energies, (see fig.2). Arousal of conformal factor, $J_{\m\n}$, is
explained below. Let us now compute the contribution of fig. 4a including the color
factors in conversion to $\<J\bar{J}\>$. Using
the Feynman rules for in fig. A1,
%%%%
\bea\la{4a}
{\mathrm{Fig\,\,4a}}&\to&-\frac{g_{YM}^2}{2}(f^{acp}f^{dhp}+f^{dcp}f^{ahp})
(-\frac{g}{2}f^{edh})(-\frac{g}{2}f^{cbe})\nonumber\\
{}&&\frac{1}{(4\pi^2)^4}\int\frac{\ud^4u}{(x-u)^2}\left(\frac{1}{(x-y)^2}\ad_{\n}^y
\frac{1}{(y-u)^2}\ad_{\m}^u\frac{1}{(x-u)^2}\right)\nonumber\\
{}&=&-N^2\delta^{ab}\frac{9}{32}\frac{g^4}{(4\pi^2)^4}\frac{1}{(x-y)^2}\ad_{\n}\p_{\m}I_2(x,y)
\nonumber
\eea
%%%%
where the integral is 
%%%%
\be\la{I2}I_2(x,y)=\int\frac{\ud^4u}{(x-u)^4}\frac{1}{(y-u)^2}=
\pi^2\frac{\ln((x-y)^2\Lambda^2)}{(x-y)^2}
\ee
%%%%
again by use of (\ref{dr1}) and (\ref{dr2}). The anomalous
contribution is obtained by keeping terms proportional to $\ln$ in
(\ref{4a}) which gives, 
$${\mathrm{Fig\,\,4a}}\to -\frac{9}{64}\frac{g^4}{4\pi^2}\d^{ab}N^2\ln((x-y)^2\Lambda^2)\frac{J_{\m\n}(x,y)}{(x-y)^2}G(x,y)^2.$$
Putting this in (\ref{totu}) together with (\ref{FSE}) one gets, 
%%%%
\bea\la{SEGE}
\<\Tr(J_{\m}\bar{J}_{\n})\>_{g.e.+s.e.}&=&-\frac{2}{g_{YM}^2}\d_{ab}
(4\times\frac{9}{64}-\frac{1}{4})\d^{ab}\frac{g_{YM}^2}{4\pi^2}N^2\nonumber\\
{}&&\ln((x-y)^2\Lambda^2)
\frac{J_{\m\n}(x,y)}{(x-y)^2}G(x,y)^2\nonumber\\
{}&=&-\frac{5g_{YM}^2N}{4\pi^2}(\frac{N}{2})^3\ln((x-y)^2\Lambda^2)\nonumber\\
{}&&\frac{J_{\m\n}(x,y)}{(x-y)^2}G(x,y)^2.
\eea
%%%%
This is the total contribution to $\<J\bar{J}\>$ from D-term, gluon
exchange and self energies. Insertion of this result into (\ref{D2pf})
yields the desired result, (\ref{andim1}).

Our next task is to fill in the details in the computation of section
3.3 that leads to the contribution of external gluons to the $\O(\l')$ anomalous
dimension, (\ref{andim2}). These contributions are shown in
fig. 6. To evaluate Graph I and II of fig. 6, we will need the
function $C_{\m}$ which was defined in (\ref{Cmu}). Recall that
external gluon is coming from the commutator $ig[A_{\m},Z]$ which
contributes $-gf^{acd}$ at the external vertex where $c$ is associated with the gluon, $d$
with $Z$-line and $a$ is the color factor of the external vertex. Use
of the Feynman rule in fig. A.1. for the internal vertex gives,       
%%%%
\bea
\mbox{\raisebox{-.45truecm}{\epsfysize=1cm\epsffile{2.7.eps}}}&=&
(-gf^{acd})(-\frac{g}{2}f^{dcb}\frac{1}{(4\pi^2)^3}
\int\frac{\ud^4u}{(x-u)^2}\left(\frac{1}{(x-u)^2}\ad_{u\,\m}
\frac{1}{(y-u)^2}\right)\nonumber\\
{}&=&-\frac{g_{YM}^2}{2}N\d^{ab}\frac{1}{(4\pi^2)^3}(-\p_{\m}^y+\half\p_{\m}^x)
I_2(x,y)\nonumber\\
{}&=&
\frac{3}{16}\frac{g_{YM}^2N}{4\pi^2}\d^{ab}\p_{\m}^y\left(\ln((x-y)^2\Lambda^2)G(x,y)\right)\nonumber
\eea
%%%%
Hence we read off 
$$C_{\m}=\frac{3}{16}\frac{g_{YM}^2}{4\pi^2}\p_{\m}^y\left(\ln((x-y)^2\Lambda^2)G(x,y)\right).$$ 
Inserting this into (\ref{exgluint}) yields the total contribution to
$\<O_{\m}^n\bar{O}_{\n}^m\>$ from Graph I,
%%%%
$$\left(\frac{N}{2}\right)^{J+2}(J+2)\d_{nm}\frac{3}{16}\frac{g_{YM}^2N}{4\pi^2}G(x,y)^J
G(x,y)\ad_{\n}\p_{\m}^y\left(\ln((x-y)^2\Lambda^2)G(x,y)\right).$$
We add to this the contribution of Graph II which is identical except a phase factor of
$q\bar{r}$ and their horizontal reflections which doubles the total answer.   
Anomalous contribution is obtained by keeping terms proportional to
$\log$ which is (\ref{comm2}) after correctly normalizing according to
(\ref{vectorop1}). Contributions of Graph III and IV (and their
horizontal reflections) are identical to above except one the factor
$1+q\bar{r}$ is replaced by $-q-\bar{r}$, hence the final answer is
(\ref{andim2}).   

\section{Trace identities and F-term contribution at $\O(\l')$}

Here, we first present the trace identities which were used throughout the
calculations and work out an example to show how to use them in
calculations of $n$-point functions. The example we choose is the
special diagram---section 4.4---which arises in D-term contribution to
torus two-point functions. As another application of the trace
identities we compute the F-term contribution to anomalous dimension 
of vector operators. 

Let us fix the convention by, 
$$\Tr(T^aT^b)=\half\d^{ab}$$ and trivially extend $SU(N)$ structure
constants, $f^{abc}$, to $U(N)$ by adding the $N\times N$ matrix
$T^0=\frac{I}{\sqrt{2N}}$. For notational simplicity let us denote all
explicit generators by their index values, \ie $T^a \to a$ and replace
explicit trace of an arbitrary matrix $M$ by $\Tr(M)\to (M)$. 

Results derived in \cite{Cvitanovic} can be
used to prove the following trace identities. 
%%%%
\bea\la{trid}
(MaM'a) &=& \half (M)(M')\qquad (ab) = \half\delta^{ab}\nonumber\\
(Ma)(aM') &=& \half (MM')\qquad (a) = \sqrt{N/2} \delta^{a0}\\
 aa &=& \half N I\qquad (~) = N\nonumber
\eea     
%%%%

%%%%%%%%%%%%%%%%%%%%%%%%%%%%%%%%%%%%%%%%%%%%%%%%%%%%%%%%%%%%%%%%%%%%%%%%%%%%%%
\begin{figure}[htb]
\centerline{ \epsfysize=6cm\epsffile{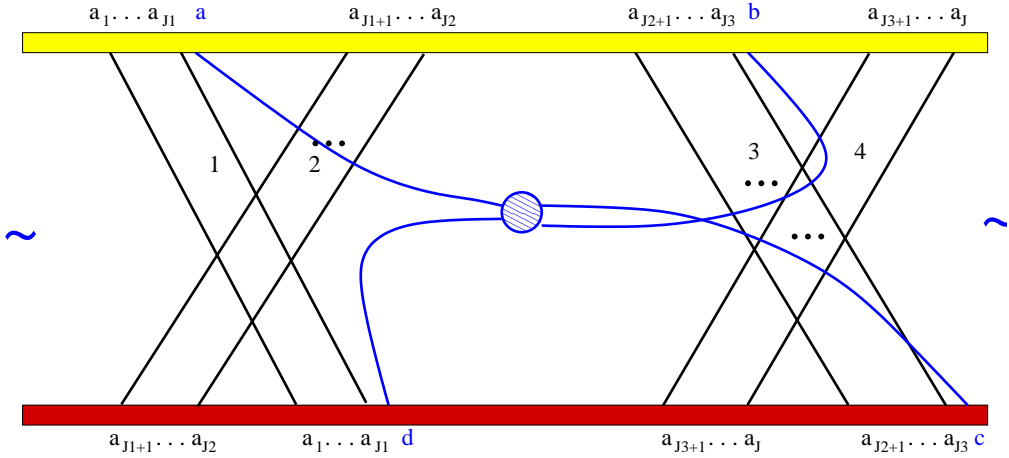}}
\caption{A torus level diagram with ``special'' topology.}
\label{B1}
\end{figure}
%%%%%%%%%%%%%%%%%%%%%%%%%%%%%%%%%%%%%%%%%%%%%%%%%%%%%%%%%%%%%%%%%%%%%%%%%%%%%%

Let us use these identities on the example of fig. 21. This shows a
D-term interaction in a special diagram. We would like to show that
the total trace involved in this graph boils down to a trace 
over the four interacting legs and eventually to show that this
diagram is indeed at torus level by algebraic methods. In fig B1, the
color indices carried by the block of $Z$-lines are $a_1,\cdots
a_{J_1}$, $a_{J_1+1},\cdots a_{J_2}$ $a_{J_2+1},\cdots a_{J_3}$ 
and $a_{J_3+1},\cdots a_{J}$ respectively. Color indices of the
interacting lines are denoted as $a,b,c,d$. 
Then the color factor associated with the vertex is $f^{acp}f^{dbp}$. 
Interestingly, untwisting the $b$ and $c$ lines in fig. 21 which give 
the other color combination, $f^{abp}f^{dcp}$ turns out to be a
genus-2 diagram! That's why we draw the special and semi-contractible 
diagrams of section 4 with twists. Use of (\ref{trid}) in fig. 21 goes
as follows, 
%%%%
\bea
&&(a_1\cdots a_{J_1} \,a\, a_{J_1+1}\cdots a_{J_2} a_{J_2+1} \cdots a_{J_3} \,b\, 
a_{J_3+1}\cdots a_{J})\nonumber\\
&&\qquad\qquad\qquad\qquad\qquad\cdot(c \,a_{J_3}\cdots a_{J_2+1} a_{J}\cdots
a_{J_3+1} \,d\, a_{J_1} \cdots a_{1} a_{J_2}\cdots a_{J_1+1})\nonumber\\ 
&=&\half(N/2)^{J-J_3-1}(a_1\cdots a_{J_1} \,a\, a_{J_1+1}\cdots a_{J_2}
a_{J_2+1} \cdots a_{J_3} \,b\, d\, a_{J_1} \cdots a_{1}a_{J_2}\cdots a_{J_1+1} 
\,c\, a_{J_3}\cdots a_{J_2+1})\nonumber\\
&=&\half(N/2)^{J-J_3-1}(b\, d\, a_{J_1} \cdots a_{1}a_{J_2}\cdots a_{J_1+1}\,c)\nonumber\\
&&\qquad\qquad\qquad\qquad\qquad \cdot (a_{J_3-1}\cdots a_{J_2+1}a_1\cdots a_{J_1} 
\,a\, a_{J_1+1}\cdots a_{J_2}a_{J_2+1} \cdots
a_{J_3-1})\nonumber\\
&=&\half(N/2)^{J-J_2-2}(b\, d\, a_{J_1} \cdots a_{1}a_{J_2}\cdots
a_{J_1+1}\,c)(a_1\cdots a_{J_1}\, a\, 
a_{J_1+1}\cdots a_{J_2})\nonumber\\
&=&\half(N/2)^{J-J_2+J_1-3}(a_{J_2}\cdots a_{J_1+1} cbda a_{J_1+1}\cdots a_{J_2})
\nonumber\\&=&(N/2)^{J-3}(cbda).\nonumber
\eea
%%%%
Contraction with the vertex color, $f^{acp}f^{dbp}$ gives
$(N/2)^{J+1}$. 

Same interaction in a planar diagram, fig. 22 would give,
%%%%
$$
(a_1\cdots a_{J_1} \,a\,b\, a_{J_1+1}\cdots a_{J}) 
(a_{J}\cdots a_{J_1+1}\,c\,d\, a_{J_1}\cdots a_{1})=\half(N/2)^{J-1}(abcd)
$$
%%%%
Contraction with the color factor of the vertex, $f^{adp}f^{bcp}$
gives, $(N/2)^{J+3}$ as the final color factor. Comparison of this
result with the special graph result shows that special graph is
indeed at torus level. 

%%%%%%%%%%%%%%%%%%%%%%%%%%%%%%%%%%%%%%%%%%%%%%%%%%%%%%%%%%%%%%%%%%%%%%%%%%%%%%
\begin{figure}[htb]
\centerline{ \epsfysize=6cm\epsffile{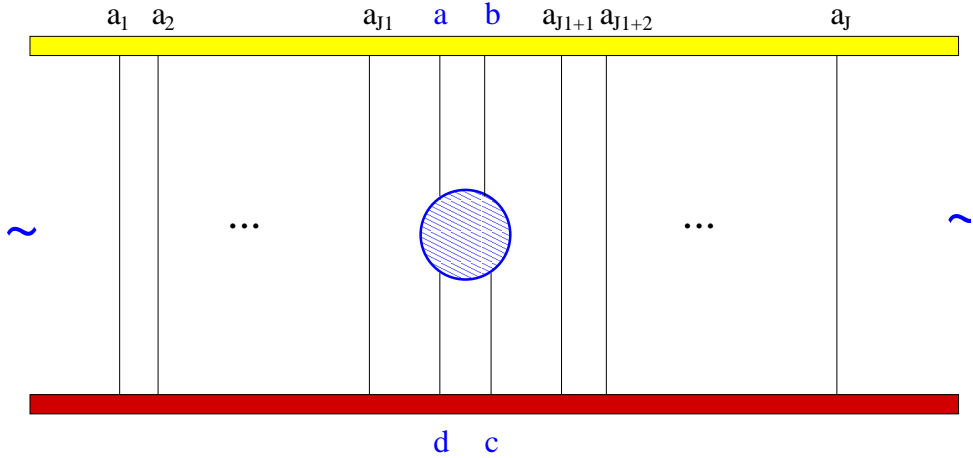}}
\caption{A planar diagram with same interaction vertex as fig. 21.}
\label{B2}
\end{figure}
%%%%%%%%%%%%%%%%%%%%%%%%%%%%%%%%%%%%%%%%%%%%%%%%%%%%%%%%%%%%%%%%%%%%%%%%%%%%%%

Let us move on to compute the F-term contribution at the planar
level. Since we are interested in the anomalous dimension we consider 
equal momenta, $n=m$ for simplicity. Unlike in the case of D-terms 
there is a nice algebraic method
to tackle with the calculation: Effective operator method \cite{Constable}. We
define the effective operator as the contraction of the vector operator 
$$O_{\m}^n=\qder(\phi Z^{J+1})=(\p_{\m}\phi Z^{J+1})+q
\sum_{l=0}^{J}q^l(\phi Z^l\p_{\m}Z Z^{J-l})$$ with
$f^{pab}\bar{Z}^a\phi^b$:
%%%%
\bea\la{oeff}
O_{eff}&=&-i\sum_{m=0}^J([a,p]Z^m a Z^{J-m})G \p_{\m}G\nonumber\\
{}&-&iq \sum_{l=0}^J q^l \sum_{m=0}^{l-1} ([a,p]Z^m a
Z^{l-m-1}\p_{\m}Z Z^{J-l})G^2\nonumber\\
{}&-&iq \sum_{l=0}^J q^l \sum_{m=0}^{J-l-1} ([a,p]Z^l a\p_{\m}Z
Z^{J-l-m-1} Z^{m})G^2\nonumber\\ 
{}&-&iq\sum_{l=0}^J q^l([a,p]Z^l a Z^{J-l})G \p_{\m}G. 
\eea   
%%%%
Planar level contribution arises from the nearest-neighbour
interactions, $m=0,J$ in the first term, $m=0$ in the second and third
terms and $l=0,J$ in the last term: 
%%%%
\be\la{oeffnn}
O'^{eff}_{\m}=i(q-\bar{q})(N/2)G\p_{\m}G(pZ^J)+iq(q-1)(N/2)G^2
\sum_{l=0}^{J-1} q^l(p Z^l \p_{\m}Z Z^{J-l-1})
\ee
%%%%
where we used (\ref{trid}) and $q^{J+2}=1$. We obtain the two-point
function as,
$\<O_{\m}^n\bar{O}_{\n}^n\>=\<O^{eff}_{\m}\bar{O}^{eff}_{\n}\>$. To
compute various terms we will need additional contraction identities, 
%%%%%%%  
\bea
\Tr(Z^a\bar{Z}^a) &=& N^{a+1} + \O(N^{a-1})
\nonumber \\
\Tr(Z^a)\Tr(\bar{Z}^a)&=& aN^{a} + \O(N^{a-2})
\label{trid2} 
\eea
%%%%%%%
which are derived by counting the number of ways one may perform the Wick 
contractions within each trace structure while obtaining a maximal power 
of $N$. Leading order terms show the planar level contributions.   

Among the four pieces in $\<O^{eff}_{\m}\bar{O}^{eff}_{\n}\>$ the term
arising from contraction of second term in $O^{eff}_{\m}$ with
second term in $\bar{O}^{eff}_{\n}$ is the easiest to evaluate. One gets,
%%%%
\be\la{cancel1}
-(q-\bar{q})^2\p_{\m}G\p_{\n}G G^{J}(N/2)^{J+3}. 
\ee
%%%%
Contraction of first term in $O^{eff}_{\m}$ with
second term in $\bar{O}^{eff}_{\n}$ gives,
%%%%
\bea\la{cancel2}
&&-(N/2)^2(q-\bar{q})q(q-1)G\p_{\m}G\sum_{l=0}^{J-1}q^l(Z^l\p_{\n}ZZ^{J-l-1}p)(p\bar{Z}^J)
\nonumber\\
&=&-(N/2)^{J+3}(q-\bar{q})q(q-1)G\p_{\m}\p_{\n}\left(\sum_{l=0}^{J-1}q^l\right)\nonumber\\
&=&(q-\bar{q})^2\p_{\m}G\p_{\n}G G^{J}(N/2)^{J+3}.\\
\eea
%%%%
The other cross-term yields the same expression hence doubles
(\ref{cancel2}). Contraction of first terms of $O^{eff}_{\m}$ and 
$\bar{O}^{eff}_{\n}$ is, 
%%%%
$$
G^4\half(N/2)^2(1-q)(1-\bar{q})\sum_{l,l'=0}^{J-1}q^l\bar{q}^{l'}
(Z^l\p_{\m}ZZ^{J-l-1}\bar{Z}^{J-l'-1}\p_{\n}\bar{Z}\bar{Z}^{l'}).
$$
%%%%
One can break up the trace into two pieces by contracting $\p_{\m}Z$
with a $\bar{Z}$ in the first group, with $\p_{\n}\bar{Z}$ or with  
a $\bar{Z}$ in the last group. First possibility gives, (up to the
factors in front of the sum)
%%%%
\bea
&&\half\p_{\m}G\sum_{p=0}^{J-l'-2}(Z^{J-l-1}\bar{Z}^p)
(\bar{Z}^{J-l'-p-2}\p_{\n}\bar{Z}\bar{Z}^{l'}Z^l)\nonumber\\
&=&\quart\p_{\m}G\p_{\n}G\sum_{p=0}^{J-l'-2}\sum_{r=0}^{l-1}
(Z^{J-l-1}\bar{Z}^p)(Z^{r}\bar{Z}^{l'})(Z^{l-r-1}\bar{Z}^{J-l'-p-2})\nonumber\\
&=&(N/2)^{J-2}\frac{N^3}{4}G^{J-2}\p_{\m}G\p_{\n}G\sum_{l<l'}q^l\bar{q}^{l'}\nonumber.
\eea
Including the factors in front, one has, 
$$(1-q)(1-\bar{q})(N/2)^{J+3}G^J\p_{\m}G\p_{\n}G\sum_{l<l'}q^l\bar{q}^{l'}.$$
whereas the second possibility gives, 
$$(1-q)(1-\bar{q})(N/2)^{J+3}G^{J+1}\p_{\m}\p_{\n}G.$$
A similar calculation shows that third possibility yields,
$$(1-q)(1-\bar{q})(N/2)^{J+3}G^J\p_{\m}G\p_{\n}G\sum_{l>l'}q^l\bar{q}^{l'},$$ 
giving all in all, 
%%%%
\be\la{cancel3}
(N/2)^{J+3}G^{J+2} \left\{J\frac{2J_{\m\n}}{(x-y)^2}
+(1+q)(1+\bar{q})\p_{\m}G\p_{\n}G\right\}.
\ee 
%%%%
Combining the various pieces we have computed, we see that
(\ref{cancel1}), (\ref{cancel2}) and the second term in
(\ref{cancel3}) cancels out, leaving us with, 
%%%%
\be\la{f4}
(N/2)^{J+3}G^{J+2}J\frac{2J_{\m\n}}{(x-y)^2}. 
\ee 
%%%%
Taking into account the integral over the interaction vertex, 
(\ref{lambdaint}), one arrives at the final contribution of the
F-terms, 
%%%%%%%%%%%%%%%%%%%%%%%%%%%%%%%%%%%%%%%%%%%%%%%%%%%%%%%%%%%%%%%%%%%%%%%%%%%%%%%
\be\la{last}
\<O_{\m}^n(x)\ol{O}_{\n}^m(y)\>_{F}=-\lambda'n^2
\ln\left((x-y)^2\Lambda^2\right)\frac{J_{\m\n}(x,y)}{(x-y)^2}G(x,y)^{J+2}
\ee
%%%%%%%%%%%%%%%%%%%%%%%%%%%%%%%%%%%%%%%%%%%%%%%%%%%%%%%%%%%%%%%%%%%%%%%%%%%%%%% 
which exactly equals the sum of the contributions from D-terms, self-energy and 
external gluons. Therefore total anomalous dimension is twice the dimension in 
\ref{last}). 

One can use the effective operator method also to calculate the F-term
contribution to torus anomalous dimension. For that purpose one should
keep the second order terms in the expansion of (\ref{trid2}). 
This calculation was computed in Appendix D of \cite{Constable} and
one gets the same expression as (\ref{torus3}).

\end{document}